\documentclass[aps,prb,reprint,groupedaddress]{revtex4-2}

\usepackage{graphicx}
\usepackage{dcolumn}
\usepackage{amsmath}
\usepackage{bm}
\usepackage{physics}
\usepackage{xcolor}

\newcommand{\comment}[1]{}

\newcommand{\ophat}[1]{\hat{#1}}
\newcommand{\ham}{\ophat{\mathcal{H}}}
\renewcommand{\phi}{\varphi}
\renewcommand{\epsilon}{\varepsilon}

\newcommand{\spin}{\mathcal{S}}
\newcommand{\opS}{\ophat{S}}
\newcommand{\pp}{\phi_k}
\newcommand{\tp}{\theta_k}
\newcommand{\tilt}{\tau}
\newcommand{\minTI}{\frac{\mu}{\lambda+\tilt}}
\newcommand{\maxTI}{\frac{\mu}{\lambda-\tilt}}
\newcommand{\minTII}{\frac{\mu}{\tilt+\lambda}}
\newcommand{\maxTII}{\frac{\mu}{\tilt-\lambda}}
\newcommand{\taul}{\tau_\lambda}
\newcommand{\muell}{\mu_\lambda}

\graphicspath{{images/}}

\begin{document}


\title{Optical conductivity of tilted higher pseudospin Dirac-Weyl cones}

\author{W. Callum Wareham}
\author{E. J. Nicol}
\email{enicol@uoguelph.ca}
\affiliation{Department of Physics, University of Guelph,
Guelph, Ontario N1G 2W1, Canada} 
\date{\today}

\begin{abstract}{
    We investigate the finite-frequency optical response of systems described at low energies by Dirac-Weyl Hamiltonians with higher pseudospin $\spin$ values. In particular, we examine the situation where a tilting term is applied in the Hamiltonian, which results in tilting of the Dirac electronic band structure. We calculate and discuss the optical conductivity for the cases $\spin=1$, $3/2$, and 2, in both two and three dimensions in order to demonstrate the expected signatures in the optical response. We examine both undertilted (type I) and overtilted (type II) as well as the critically-tilted case (type III). Along with the well-known case of $\spin =1/2$, a pattern emerges for any $\spin$. We note that in situations with multiple nested cones, such as happens for $\spin>1$, the possibility of having one cone being type I while the other is type II allows for more rich variations in the optical signature, which we will label as type IV behavior. We also comment on the presence of optical sum rules in the presence of tilting. Finally, we discuss tilting in the $\alpha$-T$_3$ model in two dimensions, which is a hybrid of the $\spin=1/2$ (honeycomb lattice) and $\spin=1$ (dice or T$_3$ lattice) model with a variable Berry's phase. We contrast this model's conductivity with that of $\spin=3/2$ and $\spin=2$ as the resultant optical response has some similarities, although there are clear distinguishing features between the these cases.
}
\end{abstract}

\maketitle

\section{Introduction}

\begin{figure}[ht]
    \includegraphics[width = \columnwidth]{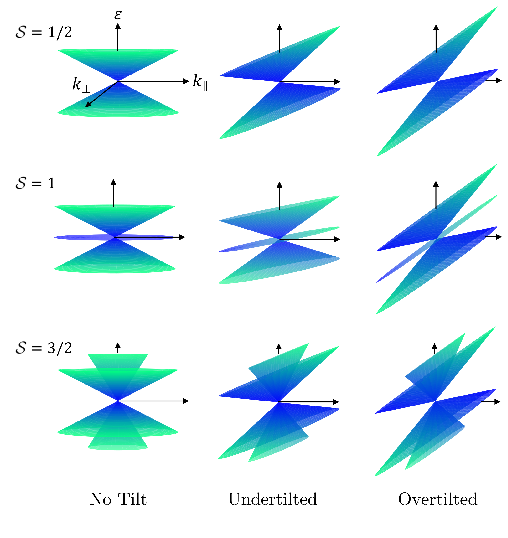}
    \caption{Schematic of energy band structures generated by Eqs. (\ref{eq:eigenvals}) and (\ref{eq:eigen_zero}) for the first three pseudospin values: $\spin = 1/2$ (row 1), $\spin = 1$ (row 2) and $\spin = 3/2$ (row 3). Untilted, undertilted, and overtilted band structures are shown in columns 1, 2, and 3 respectively.} \label{fig:cones}
\end{figure}

The experimental isolation of graphene \cite{novoselovElectricFieldEffect2004, novoselovTwodimensionalAtomicCrystals2005, castronetoElectronicPropertiesGraphene2009a, geimRiseGraphene2007} has been remarkable due to wide-reaching implications for technological applications, materials science, and fundamental physics. Relevant for this work is the impact this discovery has had on fundamental theory and developments with regard to the emerging field of Dirac-Weyl materials. Graphene provides a solid state analogue system for the Dirac equation of high-energy physics \cite{semenoffCondensedMatterSimulationThreeDimensional1984}.
This arises from the low energy physics of the electron dynamics in graphene which maps onto a Dirac equation for massless fermions with spin $\spin =1/2$. Here, however, the `spin' is actually a `pseudospin' associated with the two hexagonal (A and B) sublattices composing the honeycomb lattice structure of a single layer of carbon atoms. The energy spectrum resulting from a tight-binding calculation exhibits a linear dependence in electronic wavevector in the low-energy limit, near the corners of the first Brillouin Zone \cite{wallaceBandTheoryGraphite1947,castronetoElectronicPropertiesGraphene2009a}. This appears as gapless valence and conduction cones in the energy band structure about the charge neutrality point (Fig. \ref{fig:cones}, top left), referred to as the `Dirac cone' and `Dirac point', respectively \cite{castronetoElectronicPropertiesGraphene2009a}. In addition, the Fermi level may be varied, using voltage gating or doping to range over this special region. Both the band structure and the change in Fermi level with doping have been verified in angle-resolved photoemission \cite{bostwickQuasiparticleDynamicsGraphene2007}.
This remarkable behavior has allowed for the prediction and experimental verification of several phenomena arising from Dirac physics, for example, the unusual quantum Hall effect \cite{novoselovTwodimensionalGasMassless2005, zhangExperimentalObservationQuantum2005} and Klein tunneling \cite{youngQuantumInterferenceKlein2009, standerEvidenceKleinTunneling2009}.

The success with graphene has given rise to research into many variations of the classic two-dimensional (2D) Dirac case and beyond, including three-dimensional (3D) Weyl materials and higher pseudospin values, both of interest in this work. Possible manifestation of these variations can found in the low energy physics of more complicated materials \cite{abergelPropertiesGrapheneTheoretical2010,armitageWeylDiracSemimetals2018,hasanWeylDiracHighfold2021,doraLatticeGeneralizationDirac2011a,urbanBarrierTransmissionDiraclike2011,tangMultipleTypesTopological2017} or can potentially be designed through other systems (for e.g., Ref.~\cite{milicevic2019}).
With the shift to materials with more elaborate band structures, it has been observed that the Dirac cones may be `tilted'. In this case, the Fermi surface is an ellipse instead of a circle (type-I tilted cone) or a pair of open hyperbola or parabola Fermi surfaces (type-II or III, respectively)
with both an electron and a hole pocket. The possibility of a tunable tilt via varying a transverse electric field, say, has been suggested for \textit{8Pmmn} borophene, a material which has been argued to be described by a 2D Dirac Hamiltonian with an extra contribution that tilts the cones. This could have implications for modelling blackhole physics \cite{farajollahpourSolidstatePlatformSpacetime2019}.
The ideas surrounding tilted Dirac cones have been around for a while for both 2D and 3D cases and in a magnetic field, \cite{goerbigTiltedAnisotropicDirac2008} and the tilt parameter has been predicted to have an effect on properties such as frequency-dependent spectroscopies \cite{nishineTiltedConeInducedCusps2010,suzumuraDynamicalConductivityDirac2014a,margulisOpticalPropertiesTwodimensional2021, tanSignaturesLifshitzTransition2022, wildOpticalAbsorptionTwodimensional2022,carbotteDiracConeTilt2016a,sonowalGiantOpticalActivity2019,herreraKuboConductivityAnisotropic2019,vermaEffectElectronholeAsymmetry2017,mojarroOpticalPropertiesMassive2021}.

In this work we wish to examine the case of tilting for higher pseudospin $\spin$ Dirac-Weyls in 2D and 3D with specific attention to the optical conductivity at finite frequency as this is a spectroscopy which can be sensitive to both tilting and pseudospin $\spin$ effects.
Optical spectroscopy has already been proven as an excellent experimental probe of Dirac physics. There has been successful demonstration of agreement between theory and experiment for graphene, \cite{andoDynamicalConductivityZeroMode2002,gusyninTransportDiracQuasiparticles2006,liDiracChargeDynamics2008,nairFineStructureConstant2008,wangGateVariableOpticalTransitions2008} bilayer graphene, \cite{abergelPropertiesGrapheneTheoretical2010,nicolOpticalConductivityBilayer2008,zhangDeterminationElectronicStructure2008,kuzmenkoInfraredSpectroscopyElectronic2009} and other materials \cite[e.g.][]{timuskThreedimensionalDiracFermions2013,orlitaObservationThreedimensionalMassless2014,malcolmMagnetoopticsMasslessKane2015,Neubauer:2016}.

Untilted Dirac cones with general pseudospin has been previously examined for predictions of finite frequency conductivity in zero magnetic field by Dora et al. \cite{doraLatticeGeneralizationDirac2011a}) and in finite magnetic field by Malcolm and Nicol \cite{malcolmMagnetoopticsGeneralPseudospin2014}. The optical conductivity in a model which varies between the graphene ($\spin=1/2$) lattice to the dice or {\cal T}$_3$ ($\spin=1$) lattice, called the $\alpha$-{\cal T}$_3$ model \cite{Piechon:2015}, has also been calculated for both $B=0$, finite $B$, and for a semi-Dirac case \cite{illesHallQuantizationOptical2015,illesMagneticPropertiesEnsuremath2016,carbotteOpticalPropertiesSemiDirac2019}. Experimental work on a mercury-cadmium-tellurium (MCT) system \cite{orlitaObservationThreedimensionalMassless2014} led to the calculation of a 3D model for the conductivity of Kane fermions both in zero and finite B, and it was suggested that the low-energy physics was a 3D manifestation of the $\alpha$-{\cal T}$_3$ model for a specific $\alpha$ value relating to an unusual Berry phase \cite{malcolmMagnetoopticsMasslessKane2015}. 
Our goal in this paper is to advance this literature and to  anticipate potential future developments in materials by providing the optical fingerprints of tilted Dirac cones in higher pseudospin $\spin$ and in the $\alpha$-{\cal T}$_3$ model.

This paper is organized as follows.
In section~\ref{sec:theory}, we outline the basic theoretical approach for the starting Hamiltonian and the method of evaluation for the conductivity. We discuss the optical conductivity of untilted cones with general $\spin$ in Section~\ref{sec:untilted}. Sections~\ref{sec:2D_inter} and \ref{sec:3D_inter} describe our results for the interband conductivity of tilted 2D and 3D Dirac-Weyl cones, respectively, with a specific focus on $\spin=1/2$, 1, $3/2$, and 2. Section~\ref{sec:intra} describes our results for the intraband (Drude) part of the optical conductivity. Spectral sum rules and velocity anisotropy are discussed in Section~\ref{sec:sum_rules}, while the optical conductivity results for a tilted $\alpha$-{\cal T}$_3$ model are described in Section~\ref{sec:a-T3}. Finally, section~\ref{sec:summary} provides the summary and concluding remarks.

\section{Theoretical Formalism\label{sec:theory}}

The simplest low energy band structure of Dirac-Weyl materials with general pseudospin $\spin$ can be described by the Hamiltonian \cite{doraLatticeGeneralizationDirac2011a}
\begin{equation}
    \ham_{0} = \hbar v\vec{k}\cdot\vec{S}, \label{eq:ham_notilt}
\end{equation}
where $v$ is a velocity, $\vec{S} = \left(\opS_x, \opS_y, \opS_z\right)$ is a vector of the spin matrices associated with spin $\spin$, and $\vec{k} = (k_x, k_y, k_z)$ is the electronic wavevector ($k_z = 0$ in the case of a 2D material).
Note that for the case of graphene ($\spin=1/2$), the Hamiltonian is normally written with Pauli matrices defined with eigenvalues $\pm 1$ and in that case $v$ would be $v_F$, but here we follow Dora {\it et al.} \cite{doraLatticeGeneralizationDirac2011a}, where these are the standard spin matrices with the eigenvalues of $\opS_z$ given as $n =-\spin, -\spin+1, \ldots, \spin$. Hence, for $\spin=1/2$, this introduces an extra factor of 1/2 making $v_F=v/2$ when comparing to the graphene literature. 
Including velocity anisotropy in the Hamiltonian is straightforward and we will provide this discussion in Sec.~\ref{sec:sum_rules}.

The general Hamiltonian given above has energy eigenvalues
\begin{equation}
    \epsilon_{n}(\vec{k}) = n \hbar v k. \label{eq:notilt_eigenvals}
\end{equation}
For $n$ nonzero, this dispersion will give rise to a series of nested cones depending on $\spin$ value (see Fig.~\ref{fig:cones}).
It will be convenient for our discussion to separate out the conical bands from the flat band at zero energy that occurs in the integer spin case.
Consequently, we will use the notation of $\epsilon_{\pm\lambda}(\vec{k})=\pm\lambda\hbar vk$, where $\lambda = 1/2, 3/2,...,\spin$ for half integer spins and $1, 2,...,\spin$ for integer spins.
The flat band, which occurs for integer spin only, is denoted as $\epsilon_0(\vec{k}) = 0$.

Referring to Fig.~\ref{fig:cones} (first column), we show the pattern of energy bands for the three lowest pseudospin values.
For $\spin = 1/2$, this corresponds to a valence and a conduction Dirac cone (indexed by $n=-1/2$ and $1/2$), which touch at the charge neutrality point corresponding to chemical potential $\mu=0$.
The $\spin = 1$ case retains the Dirac cone structure of the previous case but now has a flat band at zero energy (the bands are indexed by $-1$, 0 and 1).
For $\spin=3/2$, there is now a pair of nested Dirac cones at positive and negative energy (indexed by $-3/2$, -1/2, 1/2, and 3/2). The $\spin=2$ case is not shown in Fig.~\ref{fig:cones}, but in this case there are a double set of cones like the $\spin=3/2$ case and a flat band as found for $\spin=1$ (indexed by $-2$, $-1$, 0, 1 and 2).
For higher pseudospin, this pattern repeats, with each new value of $\lambda$ contributing an additional valence and conduction cone. 

We investigate `tilted' cones in this work by adding an additional term proportional to the identity operator $\opS_0$ in the Hamiltonian, which applies a tilt to the band structure. The modified Hamiltonian is then
\begin{equation}
    \ham_{\tilt} = \hbar v (\vec{k}\cdot\vec{S} + \tilt k_\parallel\opS_0), \label{eq:basic_ham}
\end{equation}
where $k_\parallel$ defines the tilt direction (it is convenient to choose $k_\parallel = k_z$ in the 3D case, while typically $k_x$ or $k_y$ is chosen in the 2D case) and $\tilt$ is the parameter controlling the tilt.
The eigenvalues become
\begin{equation}
    \epsilon_{\pm \lambda}(\vec{k}) = \hbar v(\pm \lambda k + \tilt k_\parallel) \label{eq:eigenvals}
\end{equation}
and, in addition for the case of integer $\spin$,
\begin{equation}
    \epsilon_0(\vec{k}) = \hbar v\tilt k_\parallel \label{eq:eigen_zero}
\end{equation}
so that now each pair of eigenenergies indexed by $\lambda$ corresponds to a pair of `tilted' 2D or 3D cones, and the flat band is now a tilted 2D or 3D (hyper)-plane (see the second and third columns of Fig. \ref{fig:cones} for a schematic illustration).
When the chemical potential $\mu$ is nonzero, the Fermi surfaces will have the form of a conic section. Specifically, the undertilted case (type-I, $\tilt<\lambda$) produces an ellipse(oid), while the critically tilted (type-III, $\tilt=\lambda$) and overtilted (type-II, $\tilt>\lambda$) cases produce open parabola(oid) and hyperbola(oid) Fermi surfaces, respectively.
The plane-type band produces a Fermi surface that is either an open line or an open plane (dependent on the dimensionality).

To calculate the optical conductivity, the standard Kubo formulation is employed for the current-current correlation function.
Given $\epsilon_{\pm \lambda}(\vec{k})$, $\epsilon_0(\vec{k})$, and the associated eigenvectors $\ket{\pm \lambda}$ and $\ket{0}$, the real part of the longitudinal optical conductivity $\sigma(\omega)$ can be computed from
\begin{widetext}
  \begin{equation}
    \sigma_{ii}(\omega) = \hbar \sum_{n,n'} \sum_{\vec{k},\vec{k'}}\frac{\Theta\left[\mu - \epsilon_{n'}(\vec{k}')\right] - \Theta\left[\mu - \epsilon_n(\vec{k})\right]}{\epsilon_n(\vec{k})-\epsilon_{n'}(\vec{k}')} 
    \times\pi \delta\left[\hbar\omega + \epsilon_n(\vec{k})-\epsilon_{n'}(\vec{k}')\right]\mel{n}{\ophat{j}_i}{n'}\mel{n'}{\ophat{j}_i}{n}, \label{eq:kubo_wavefunction}
    \end{equation}
\end{widetext}
where $n$ and $n'$ index over the allowed band indices and $\omega$ is the photon frequency. The chemical potential $\mu$ is taken to be positive throughout this work, for notational simplicity.
$\Theta(x)$ and $\delta(x)$ are the Heaviside step and Dirac delta functions, respectively, and $\ophat{j}_{i} = -\frac{e}{\hbar}\frac{d\ham}{dk_i}=-e\ophat{v}_i$ are the current operators.
We have taken the limit of zero-temperature ($T=0$) and zero impurity scattering in order to capture the essential physics.
Finite temperature and impurity scattering serve to broaden and smooth out sharp edges on  the curves.
Here, we will solely discuss the real (absorptive) part of the conductivity; 
the imaginary part can be obtained through Kramers-Kronig evaluation.

Finally, note that throughout this work we present the optical conductivity results for a single electron spin and valley.
Spin and valley degeneracy can be restored by multiplying our results by a factor $g_s$ and $g_v$, respectively.

As an illustrative example, the Hamiltonian, wavefunctions and relevant matrix elements for the tilted 3D $\spin = 3/2$ case are presented below.
Working in spherical coordinates ($\tan\pp = k_y/k_x$ and $\cos\tp = k_z/k$), the Hamiltonian is
\begin{widetext}
\begin{equation}
    \ham_{\tilt,~3/2} = \hbar v k\begin{pmatrix}
        (\tilt + \frac{3}{2})\cos\tp  & \frac{\sqrt{3}}{2}e^{-i\pp}\sin\tp & 0 & 0 \\
        \frac{\sqrt{3}}{2}e^{i\pp}\sin\tp & (\tilt + \frac{1}{2})\cos\tp & e^{-i\pp}\sin\tp & 0 \\
        0 & e^{i\pp}\sin\tp & (\tilt -\frac{1}{2})\cos\tp  & \frac{\sqrt{3}}{2}e^{-i\pp}\sin\tp \\
        0 & 0 & \frac{\sqrt{3}}{2}e^{i\pp}\sin\tp & (\tilt-\frac{3}{2})\cos\tp
    \end{pmatrix}. \label{eq:ham_32}
\end{equation}
The eigenvalues are as given by Eq.~(\ref{eq:eigenvals}), and the eigenvectors are
\begin{subequations}
\begin{equation}
    \ket{\pm1/2} = N_{\pm1/2}\begin{pmatrix}
                e^{-3i\pp}\sin \tp/(\cos \tp\mp1) \\
                \pm\sqrt{3}e^{-2i\pp}(\frac{1}{3}\mp\cos\tp)/(\cos\tp \mp 1) \\
                \sqrt{3}e^{-i\pp}(\cos\tp \pm \frac{1}{3})/\sin\tp \\
                1
                \end{pmatrix}, 
\end{equation}
\begin{equation}
    \ket{\pm3/2} = N_{\pm3/2}\begin{pmatrix}
        \pm e^{-3i\pp}\tan^{\mp3}(\tp/2) \\
        \sqrt{3}e^{-2i\pp} \tan^{\mp2}(\tp/2) \\
        \pm \sqrt{3}e^{-i\pp} \tan^{\mp1}(\tp/2) \\
        1
        \end{pmatrix},
\end{equation}
\end{subequations}
\end{widetext}
with each $N_{\pm\lambda}$ denoting the normalization factors such that each $\braket{\pm\lambda}{\pm\lambda}=1$.
Notice that all these eigenvectors are $\tilt$-independent and are therefore the same in the untilted limit 
(this is unsurprising, given that the untilted and tilted Hamiltonians commute with each other and should therefore share eigenvectors).
Using these eigenvectors and the velocity operators $\ophat{v}_i$, the relevant non-zero matrix elements entering Eq.~(\ref{eq:kubo_wavefunction}) for $\sigma_{zz}$ (the tilted direction) are
\begin{equation}
    |\mel{\pm \lambda}{\ophat{v}_z}{\pm \lambda}|^2 = (\tilt\pm\lambda \cos\tp)^2,
\end{equation}
\begin{equation}\displaystyle
    |\mel{\pm 1/2}{\ophat{v}_z}{\pm 3/2}|^2 = \frac{3}{4}\sin^2\tp,
\end{equation}
\begin{equation}
    |\mel{\pm 1/2}{\ophat{v}_z}{\mp 1/2}|^2 = \sin^2\tp,
\end{equation}
and for $\sigma_{xx}$ (untilted direction),
\begin{equation}
    |\mel{\pm \lambda}{\ophat{v}_x}{\pm \lambda}|^2 = \lambda^2\sin^2\tp\cos^2\pp,
\end{equation}
\begin{equation}\displaystyle
  |\mel{\pm 1/2}{\ophat{v}_x}{\mp 1/2}|^2 = \frac{1}{4}[3+\cos(2\tp)-2\sin^2\tp\cos(2\pp)],
\end{equation}
\begin{equation}\displaystyle
  |\mel{\pm 1/2}{\ophat{v}_x}{\pm 3/2}|^2 = \frac{3}{4}  |\mel{\pm 1/2}{\ophat{v}_x}{\mp 1/2}|^2.\label{eq:last_el_32}
\end{equation}
The matrix elements and the energy eigenvalues enter the formula for the optical conductivity.

\section{Untilted Results} \label{sec:untilted}


To better appreciate the effect of the tilt for higher pseudospin, we first examine the optical conductivity in the untilted case, {\it i.e.} $\tilt=0$.
The 2D untilted case has been calculated for a general $\spin$ by D\'ora \textit{et al.} \cite{doraLatticeGeneralizationDirac2011a}. Here, we review the 2D results and provide our results for 3D.

At finite $\mu$, there are two components to the optical conductivity: the Drude (or intraband) component centered around zero frequency, and the interband contribution at finite frequency. 
The intraband response is associated with transitions within an energy band right near the Fermi level. It may be broadened into a Lorentzian form with impurity scattering (the Drude form) or simply be a delta function at $\omega=0$ in the case of the pure limit.
The interband conductivity is due to the transitions between energy bands allowed by selection rules.
In this paper, we will primarily focus on the interband form as providing the most evident features in spectroscopy, but we will also provide results for the intraband component and discuss the possibility of sum rules or spectral weight transfer between the two components.

The result for pseudospin $\spin$ in 2D from Dora \textit{et al.} \cite{doraLatticeGeneralizationDirac2011a} is rewritten here for $T=0$ and with notation to match our paper. The interband conductivity (per valley and electron spin) is
\begin{equation}
  \sigma^{2D}_{\rm inter}(\omega) = \sigma_0\sum_{\lambda = \lambda_{\rm min}}^\spin \lambda\theta(\lambda\hbar\omega-\mu),
  \label{eq:2D_inter_notilt}
\end{equation}
and the intraband part is
\begin{equation}
        \sigma^{2D}_{\rm intra}(\omega) = 2\mu\sigma_0 \lfloor \spin + 1/2 \rfloor \delta(\omega), \label{eq:2D_intra_notilt}
\end{equation}
where $\sigma_0 = e^2/8\hbar$,
$\lambda_{\rm min} = 1/2$ or $1$ for either half-integer or integer $\spin$, respectively, and $\lfloor x \rfloor$ is the floor function. To recover the result for graphene, which is $\spin=1/2$, a factor of four for the valley and electron spin degeneracy must be included, and then
the conductivity becomes the well-known form of $\sigma(\omega)=2\sigma_{0}[4\mu\delta(\omega)+\theta(\hbar\omega-2\mu)]$.

In the 3D case from our calculations up to $\spin = 2$, using the conductivity formula in Eq.~(\ref{eq:kubo_wavefunction}), we find the interband and intraband parts of the conductivity are
\begin{equation}
    \sigma^{3D}_{\rm inter}(\omega) = A\omega\sum_{\lambda=\lambda_{\rm min}}^\spin \lambda\theta(\lambda\hbar\omega-\mu), \label{eq:3D_inter_notilt}
\end{equation}
and
\begin{equation}
    \sigma^{3D}_{\rm intra}(\omega) = \delta(\omega) \mu^2A\sum_{\lambda = \lambda_{\rm min}}^\spin 1 /\lambda, \label{eq:3D_intra_notilt}
\end{equation}
where $A = e^2/(12\pi\hbar v)$. 
This form differs from the 2D case primarily by the effect of the electronic density of states, which in the higher dimension provides an extra factor of $\omega$.

\begin{figure}[ht]
    \includegraphics[width = \columnwidth]{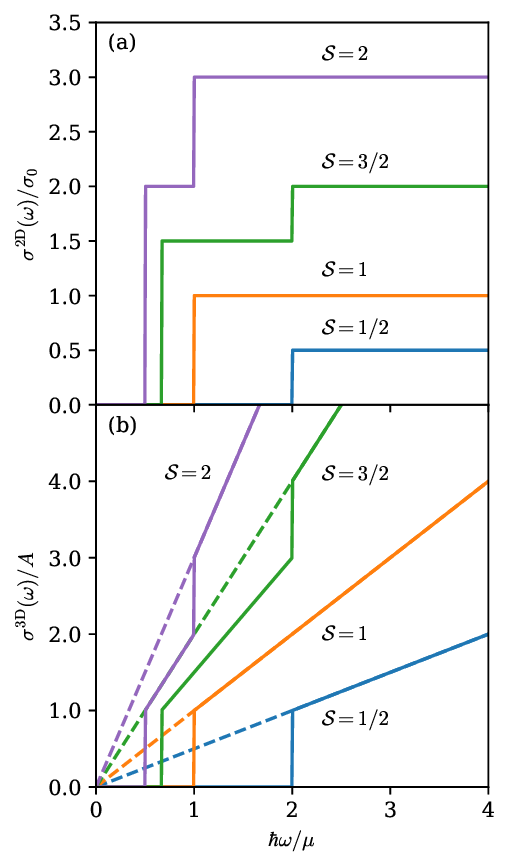}
    \caption{Interband optical conductivity due to untilted pseudospin-$\spin$ cones in the (a) 2D \cite{doraLatticeGeneralizationDirac2011a}
      and (b) 3D cases. The vertical dashed grey lines at $\hbar \omega/\mu = 1$ and  $\hbar \omega/\mu = 2$ represent the absorption steps generated by  $\lambda = 1$ and $\lambda = 1/2$ cones, respectively; additional cones for integer or half-integer spins produce absorption steps at $\hbar\omega/\mu = 1/\lambda$ which are all below $\hbar\omega/\mu=1$.\label{fig:2D_3D_untilted}}
\end{figure}

\begin{figure}[ht]
    \includegraphics[width=\columnwidth]{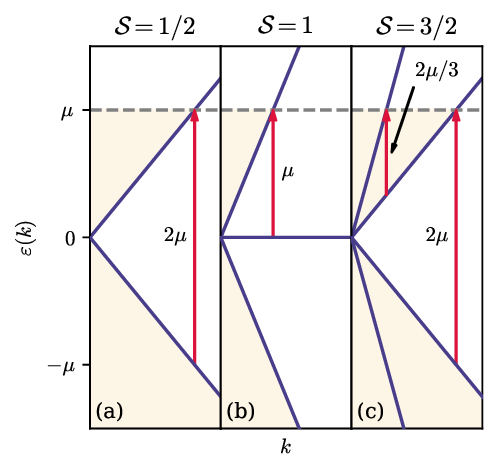}
    \caption{Schematic of smallest allowed transitions (red arrows) between bands (blue lines) in the untilted case for (a) $\spin = 1/2$, (b) $\spin = 1$, and (c) $\spin = 3/2$. Shaded regions represent filled electron states up to the chemical potential $\mu$ (grey dashed line). Notice that the smallest unblocked transitions occur at $2\mu$ and $\mu$ in the $\spin=1/2$ and $\spin=1$ cases, respectively, and at $2\mu/3$ and $2\mu$ in the $\spin = 3/2$ case. Here, the vertical axis is the energy scale of the bands, and the horizontal axis is the electron wavenumber $k$ in arbitrary units.}\label{fig:arrows_notilt}
\end{figure}

In Fig.~\ref{fig:2D_3D_untilted}, we plot the 2D and 3D interband results in the upper and lower frames, respectively, to illustrate the trend in behavior.
Aside from a flat background conductivity in 2D versus a linear background in 3D, which is due to the unique character of the linear energy bands in Dirac-Weyl systems, the major features are the various steps seen in the conductivity depending on the specific pseudospin value.
To best understand these structures, we first discuss the 2D results and rehearse the case of $\spin=1/2$ (Fig.~\ref{fig:2D_3D_untilted}(a) blue curve), where interband transitions occur between the valence band and the conduction band. If the chemical potential is non-zero then there will be filled states in the conduction band below the chemical potential.
Absorption can only occur if the energy of an incident photon is sufficient to promote an electron from an occupied state in the valence band to an unoccupied state in the conduction band.
As the photon is essentially a $q\sim 0$ probe for this experimental property, there is no momentum transfer from the photon and hence the transitions are vertical in the band structure (see Fig. \ref{fig:arrows_notilt}(a)).
The first transitions that can occur will be for a photon energy of $2\mu$.
Before this point, all possible transitions from the lower cone to the upper cone are blocked by the Pauli exclusion principle, i.e. lower-energy photons cannot promote an electron from the lower band to the upper band because there is no available final state for the electron in the upper band.
This is the physics behind the absorption step in the conductivity at $2\mu$.
Throughout this work, we assume $\mu$ is positive for simplicity, but the results also hold for negative $\mu$ if its absolute value is used in the presented equations.
Finally, for $\mu \to 0$, the step vanishes and we are left with constant absorption $\sigma_0/2$ for all values of $\hbar \omega$.

Continuing to the $\spin = 1$ case (orange curve), a similar absorption step at $\hbar \omega = \mu = \mu/\lambda$ to the constant value $\sigma = \sigma_0\lambda= \sigma_0$ is present ($\lambda = 1$ cones and a flat band form the band structure here).
The energy of this step now corresponds to the energy required to transition from the flat band to the upper cone at the Fermi level, while all lower-energy transitions are Pauli-blocked [Fig. \ref{fig:arrows_notilt}(b)].
Note that there is no additional transition step between the cones themselves (which would appear at $\hbar \omega = 2\mu$) illustrating the selection rule that electrons may only transition between neighboring bands.

To further illustrate the pattern, the $\spin = 3/2$ case (green curve) shows absorption steps at $\hbar \omega = 2\mu/3$ and $\hbar \omega = 2\mu$, or at $\hbar \omega = \mu/\lambda$ for the allowed values $\lambda = 3/2$ and $\lambda = 1/2$, respectively.
The second step is the same transition between $\lambda = 1/2$ cones as in the $\spin = 1/2$ case, while the first step corresponds to an additional transition from the upper $\lambda = 1/2$ cone to the upper $\lambda = 3/2$ cone (Fig. \ref{fig:arrows_notilt}(c)).  
The $\spin = 2$ case (violet curve) has two upper cones, two lower cones, and a flat band. Transitions will occur from the flat band to the $\lambda=1$ cone just as seen in Fig.~\ref{fig:arrows_notilt}(b) for $\spin=1$ and so there is an onset of absorption at $\hbar\omega=\mu$. An additional channel for absorption opens for transitions from the $\lambda=1$ cone to the $\lambda=2$ cone analogous to the upper cone-to-cone transitions for $\spin = 3/2$. This will give rise to a step at $\hbar\omega=\mu/\lambda=\mu/2$.
It clear that multiple steps can occur because an extra channel of absorption will result with each additional band in the band structure.
The pattern emerging is that the number of steps indicates the number of energy bands crossing the Fermi level at finite $\mu$. 
Note that the overall constant background for $\hbar\omega>\mu$ (for integer $\spin$) and $\hbar\omega>2\mu$ (for half-integer $\spin$) is given by
$\sigma_0\spin(\spin+1)/2$ and $\sigma_0[4\spin(\spin+1)+1]/8$, respectively, and is equivalent to the $\mu\to0$ limit \cite{doraLatticeGeneralizationDirac2011a}.

While the discussion here may seem somewhat theoretical, it is worthwhile to imagine the possibility of using a higher pseudospin system (such as 3/2) for a type of practical device where variation of incident light is used to to switch the conductivity of a device from ``off'' to ``on'', and the ``on'' state could have multiple levels of conduction.  

Much of the discussion above also applies to Fig. \ref{fig:2D_3D_untilted}(b) showing the 3D case up to $\spin = 2$.
Just like the 2D case, steps appear at each energy $\hbar \omega = \mu/\lambda$, and each transition now contributes a $\lambda$-dependent slope (instead of constant absorption) above the step due to the 3D density of states. 
The dashed lines in Fig. \ref{fig:2D_3D_untilted}(b) represent the $\mu=0$ curves and result from the saturated slope from all transitions together at each $\spin$.
If we examine the case of $\spin=2$ (violet curve), as an example, we note that with finite $\mu$, the two sections of the curve have different slopes, but that both linear sections will extrapolate to the origin.
This is an important point as there are experimental data in the literature, for a number of systems, where the data show linear curves but with intercepts that are negative (e.g., Refs.~\cite{timuskThreedimensionalDiracFermions2013,Neubauer:2016}).
It has been a source of research to understand this effect, which cannot be generated easily in any model, although in a tilted cone model, as we discuss here, such behavior can be shown to occur.
It is important to have this untilted 3D case for reference in what is to be presented later in this paper.

\section{Tilted 2D Interband Results\label{sec:2D_inter}}
\label{sec:2Dtilted}


Having set the foundation for the fingerprints expected in the optical conductivity in 2D and 3D, we now turn to the case of a 2D tilted band structure, focusing on the interband results.
Using Eq.~(\ref{eq:kubo_wavefunction}), we have evaluated the optical conductivity for tilted 2D $\spin = 1/2,$ $1$, $3/2$, and $2$ and find the general pattern is given by
\begin{equation}
    \sigma^{\textrm{2D}}_{i}(\omega)\ = \sigma_0\sum_{\lambda = \lambda_{min}}^\spin  \lambda\mathcal{F}^{\textrm{2D}}_{i}(\lambda,\omega), \label{eq:2D_fullinter}
\end{equation}
where $i = (\parallel, \bot)$ refers to a photon polarization parallel or perpendicular to the tilt direction, respectively, and $\mathcal{F}^{\textrm{2D}}_{i}(\lambda,\omega)$ is $\tilt$-dependent and given below for the different types of tilting.
Following what
others have done for $\spin = 1/2$ \cite{tanSignaturesLifshitzTransition2022, wildOpticalAbsorptionTwodimensional2022, houEffectsSpatialDimensionality2023}, we define
\begin{equation}
    G_\pm(x) = \frac{1}{\pi}\left(\arcsin(x) \pm x\sqrt{1-x^2}\right),
\end{equation}
and
\begin{equation}
    \xi_{\pm\lambda} = \frac{\mu \pm \hbar\omega\lambda}{\hbar\omega\tilt}.
\end{equation}
Then
for $\tilt<\lambda$ (`undertilted', type-I cones),
\begin{subequations}
\begin{equation}
    \mathcal{F}^{\textrm{2D}}_{\parallel}(\lambda,\omega) =  \begin{cases}
        0 & \hbar\omega \leq  \minTI \\
        \frac{1}{2} - G_+(\xi_{-\lambda})  & \minTI < \hbar\omega \leq \maxTI \\
        1 & \maxTI < \hbar\omega,
    \end{cases}
\end{equation}\label{eq:inter_parallel_2_under}
\begin{equation}
    \mathcal{F}^{\textrm{2D}}_{\bot}(\lambda,\omega) =  \begin{cases}
        0 & \hbar\omega \leq  \minTI \\
        \frac{1}{2} - G_-(\xi_{-\lambda})  & \minTI < \hbar\omega \leq \maxTI \\
        1 & \maxTI < \hbar\omega.
    \end{cases}
\end{equation}\label{eq:inter_perp_2_under}
\end{subequations}
For $\tilt>\lambda$, (`overtilted', type-II cones), we have
\begin{subequations}
\begin{equation}
    \mathcal{F}^{\textrm{2D}}_{\parallel}(\lambda,\omega) =  \begin{cases}
        0 & \hbar\omega \leq  \minTII \\
        \frac{1}{2} - G_+(\xi_{-\lambda})  & \minTII < \hbar\omega \leq \maxTII \\
        G_+(\xi_{+\lambda}) - G_+(\xi_{-\lambda})  & \maxTII < \hbar\omega,
    \end{cases} 
\end{equation}\label{eq:inter_parallel_2_over}
\begin{equation}
    \mathcal{F}^{\textrm{2D}}_{\bot}(\lambda,\omega) =  \begin{cases}
        0 & \hbar\omega \leq  \minTII \\
        \frac{1}{2} - G_-(\xi_{-\lambda})  & \minTII < \hbar\omega \leq \maxTII \\
        G_-(\xi_{+\lambda}) - G_-(\xi_{-\lambda})  & \maxTII < \hbar\omega.
    \end{cases} 
\end{equation}\label{eq:inter_perp_2_over}
\end{subequations}
Finally, for $\lambda=\tilt$, (`critically tilted', type-III cones), we obtain
\begin{subequations}
\begin{equation}
    \mathcal{F}^{\textrm{2D}}_{\parallel}(\lambda,\omega) =  \begin{cases}
        0 & \hbar\omega \leq  \frac{\mu}{2\lambda} \\
        \frac{1}{2} - G_+(\xi_{-\lambda})  & \frac{\mu}{2\lambda} < \hbar\omega,
    \end{cases}
\end{equation}
\begin{equation}
    \mathcal{F}^{\textrm{2D}}_{\bot}(\lambda,\omega) =  \begin{cases}
        0 & \hbar\omega \leq  \frac{\mu}{2\lambda} \\
        \frac{1}{2} - G_-(\xi_{-\lambda})  & \frac{\mu}{2\lambda} < \hbar\omega.
    \end{cases}
\end{equation}\label{eq:inter_perp_2_crit}
\end{subequations}
Setting $\lambda = 1/2$ in the above equations reproduces the well-known tilted $\spin = 1/2$ (`tilted graphene') case.
While the transitions for $\spin = 1/2$ are between the two bands indexed by $\lambda = 1/2$ ({\it i.e.}, $\epsilon_{-1/2}$ and $\epsilon_{+1/2}$) it is important to note that, for higher $\spin$ values, transitions occur between neighboring bands.
In the $\spin = 3/2$ case, for example, transitions between the $\epsilon_{+1/2}$ and $\epsilon_{+3/2}$ bands account for the $\mathcal{F}^{\textrm{2D}}_i(\lambda = 3/2,\omega)$ contribution.

\begin{figure}[ht]
    \includegraphics[width=\columnwidth]{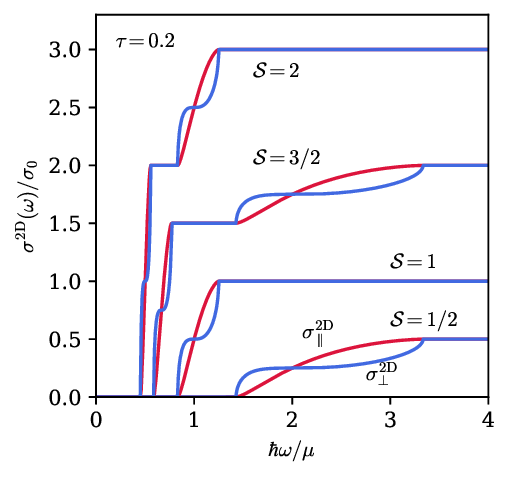}
    \caption{Frequency-dependent interband optical conductivity for slightly tilted (type-I) 2D Dirac materials with pseudospin $\spin = 1/2$, 1, $3/2$, and 2, where we have set $\tilt = 0.2$. We present the conductivity parallel ($\sigma_\parallel^{2D}$, red) and perpendicular ($\sigma_\bot^{2D}$, blue) to the direction of the tilt. Notice that each step in the $\tilt = 0$ case (Fig. \ref{fig:2D_3D_untilted}) has been broadened in a directional-dependent manner.}\label{fig:slight_tilt2d}
\end{figure}

\begin{figure}[ht]
    \includegraphics[width=\columnwidth]{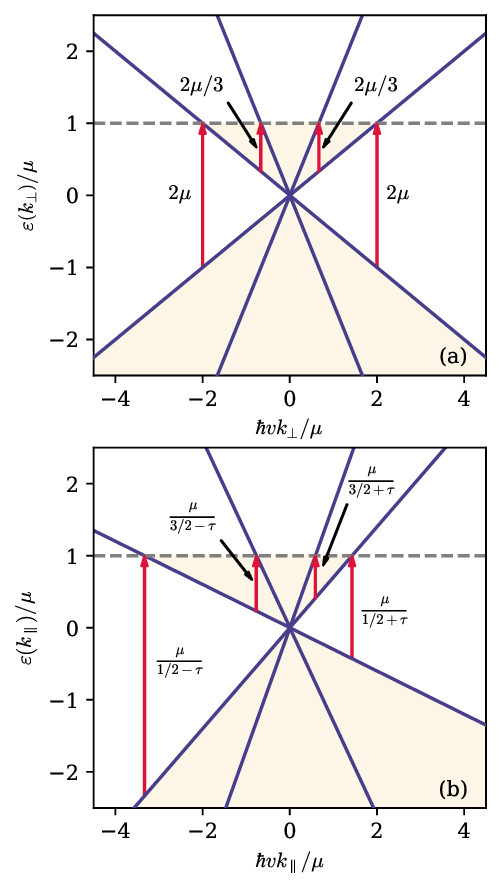}
    \caption{Schematic diagram of photon transition energies between bands (blue lines) to the Fermi level (grey dashed line) in the tilted $\spin = 3/2$ case, plotted in (a) the $k_\bot$ plane, and (b) the $k_\parallel$ plane. Transition energies are denoted by red arrows and labelled with their length. Note that transitions occur in the $k_\parallel$ plane with lengths $\mu/(\lambda\pm \tilt)$, while the band structure in the $k_\bot$ plane appears the same as the untilted case (and these transitions are therefore unaffected by the tilt).}\label{fig:arrows_tilt}
\end{figure}

The final results evaluated from Eq.~(\ref{eq:2D_fullinter}) for `undertilted' 
$\spin = 1/2$, 1, $3/2$ and 2 systems (\textit{i.e.}, $\tilt < \lambda$ for all $\lambda$) are presented in Fig. \ref{fig:slight_tilt2d}.
Notice that each step appearing in the $\tilt = 0$ case of Fig.~\ref{fig:2D_3D_untilted}(a) (at allowed values of $\hbar\omega = \mu/\lambda$) is now split into two key energy scales at $\hbar\omega = \frac{\mu}{\lambda+\tilt}$ and $\frac{\mu}{\lambda-\tilt}$.
For undertilted cones, these two scales correspond to the energies of the smallest and largest vertical transitions which end at the Fermi energy $\mu$.
Fig. \ref{fig:arrows_tilt} shows a schematic diagram of these transitions.
Along the $k_\bot$ axis [Fig. \ref{fig:arrows_tilt}(a)], the band structure is unaffected by the tilt; the $k_\parallel$ axis [Fig. \ref{fig:arrows_tilt}(b)], however, reveals the key energy scales in the optical conductivity spectrum.
For photon energies below $\frac{\mu}{\lambda+\tilt}$, all transitions are Pauli blocked, while above $\frac{\mu}{\lambda-\tilt}$, all transitions are available and the contribution to the optical conductivity from those transitions saturates at the untilted value $\lambda\sigma_0$.
Between these key energies, the conductivity is polarization (direction) dependent.
Wild et al. \cite{wildOpticalAbsorptionTwodimensional2022} explain this directional dependence in the $\spin = 1/2$ case via momentum alignment and the elliptical shape of the Fermi surface; we provide a similar phrasing of this argument below, generalizing to the higher pseudospin cases.

\begin{figure}[ht]
    \includegraphics[width=\columnwidth]{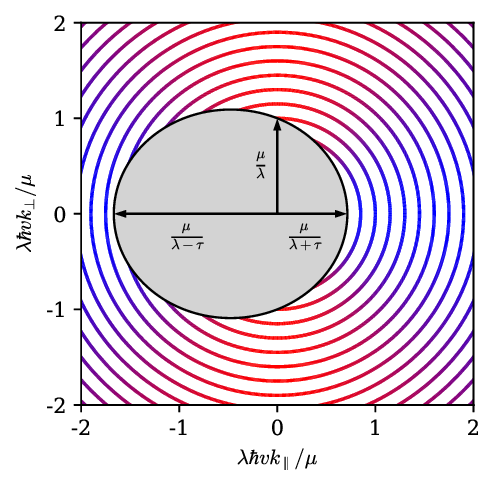}
    \caption{Schematic diagram of the Pauli-blocked transitions (grey shaded area) for an undertilted cone indexed by $\lambda$, plotted on top of concentric (constant transition energy) circles. Regions of the circles are color-coded red or blue to represent transitions which contribute strongly to $\sigma_\parallel$ or $\sigma_\bot$, respectively. `Blue' ($\sigma_\bot$-dominant) transitions are unblocked first at $\hbar\omega = \mu/(\lambda + \tilt)$, followed by `red' ($\sigma_\parallel$-dominant) transitions near $\mu/\lambda$ (at which point half of the possible transitions are unblocked), while all transitions are finally freed at $\hbar\omega = \mu/(\lambda - \tilt)$.} \label{fig:circles}
\end{figure}

Four facts assist with qualitatively understanding the shape of the optical conductivity curves.
Firstly, the energy difference between neighboring bands is always $|\epsilon_n - \epsilon_{n\pm1}| = \hbar v k$ irrespective of the tilt $\tilt$.
This means that one can draw concentric circles of electron transitions (indexed by a constant value of $k$) which all have the same vertical energy distance to the next-neighboring cone (see Fig.~\ref{fig:circles}). Although each transition may differ from the others as to its starting and end points in the band structure, each contributes to the optical conductivity $\sigma(\omega)$ at the same specific photon frequency $\omega$.
It is important to emphasize that these circles do not refer to a surface of equal electron energy on a particular band, but rather to a collection of (blocked or unblocked) vertical transitions corresponding to the same photon frequency.
Secondly, the velocity matrix element between the cones $\mel{n}{\ophat{v}_\alpha}{n\pm1}$ is $\tilt$-independent, meaning that the tilt does not affect the wavefunction-overlap part of the Kubo formula [Eq.~(\ref{eq:kubo_wavefunction})].
Thirdly, a constant optical response with increasing photon energy implies that these concentric circles of potential equal-energy transitions each contribute the same total optical response when they are fully unblocked.
Finally, electrons in Dirac-Weyl systems respond for interband transitions most strongly to photons polarized perpendicularly to their wavevector. 
In particular, electrons on the $k_\parallel$ axis only contribute to $\sigma_\bot$, and vice versa.

Taken together, these facts imply that `how much' of a concentric transition circle is unblocked, and `where' in $k$-space the unblocked parts of the circle are, can explain the directional-dependent shape of the optical conductivity curves. 
Fig. \ref{fig:circles} is a schematic diagram showing a discretized set of these transition circles, which are color-coded to represent transitions which contribute strongly to $\sigma_\bot$ (blue) or $\sigma_\parallel$ (red) corresponding to their colors in the optical conductivity curves; superimposed on these circles are Pauli-blocked transitions (grey shaded area), with the black outline representing the Fermi surface.
(It is useful to focus on the $\spin = 1/2$ case in Fig.~\ref{fig:slight_tilt2d} when connecting this schematic to the optical conductivity curves, where the distinctive features of each polarization are well-resolved.)
The circles in Fig.~\ref{fig:circles} with a radius below $\frac{\mu}{\lambda + \tilt}$ are fully blocked, and there are therefore no transitions for photon energies below this limit.
Just above this radius, only `blue' ($\sigma_\bot$-contributing) states are unblocked, corresponding to the strong jump towards half of the saturated response in $\sigma_\bot$ and essentially no response in $\sigma_\parallel$ near this point.
As the circles approach a radius of $\mu/\lambda$, half of the `red' ($\sigma_\parallel$-contributing) transitions also become unblocked; exactly half of the `blue' and half of the `red' portions are free, and the two polarization directions therefore have an equivalent response at this point.
Moving past $\mu/\lambda$, red transitions continue to be freed while the blue transitions to the left in Fig.~\ref{fig:circles} are still blocked ($\sigma_\bot$ lags behind $\sigma_\parallel$) until we approach $\frac{\mu}{\lambda - \tilt}$.
At this radius, all red transitions have been activated ($\sigma_\parallel$ is nearly constant at its saturated value) while the last of the blue transitions are quickly freed ($\sigma_\bot$ has a very large slope and quickly catches up to its saturated value).
Above this radius, all transitions are unblocked and both polarization directions yield their saturated, untilted value.

Returning to Fig.~\ref{fig:slight_tilt2d}, we note that $\spin=1/2$ and $\spin=1$ have similar shapes, however, they do not scale on to each other. A factor of 2 may be used to adjust the crossing points at $\hbar\omega=\mu$ and $2\mu$ to be the same, and to scale the saturated backgrounds to match, but the broadening of the $\spin=1/2$ curve is larger than the $\spin=1$ curve. This is a distinguishing feature between the two $\spin$ values which may be relevant should insufficient information be available about $\mu$ or the saturated background scale. Likewise in a comparison of $\spin=3/2$ with $\spin=2$, the pair of curves cannot be scaled on top of each other.

\begin{figure}[ht]
    \includegraphics[width=\columnwidth]{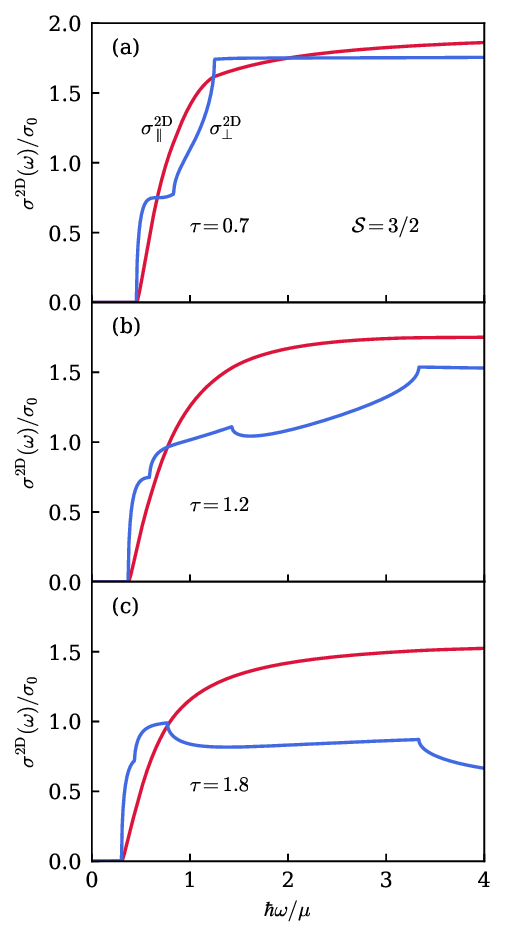}
    \caption{Interband optical conductivity of type~IV overtilted 2D $\spin=3/2$ Dirac materials, illustrating some potential optical fingerprints. For $1/2<\tilt<3/2$ [e.g. (a) $\tilt = 0.7$ and (b) $\tilt = 1.2$], only the $\lambda = 1/2$ set of cones is overtilted. For $\tilt>3/2$ [e.g. (c) $\tilt = 1.8$], however, both sets are overtilted.}\label{fig:2D_32tilts}
\end{figure}

\begin{figure}[ht]
    \includegraphics[width=\columnwidth]{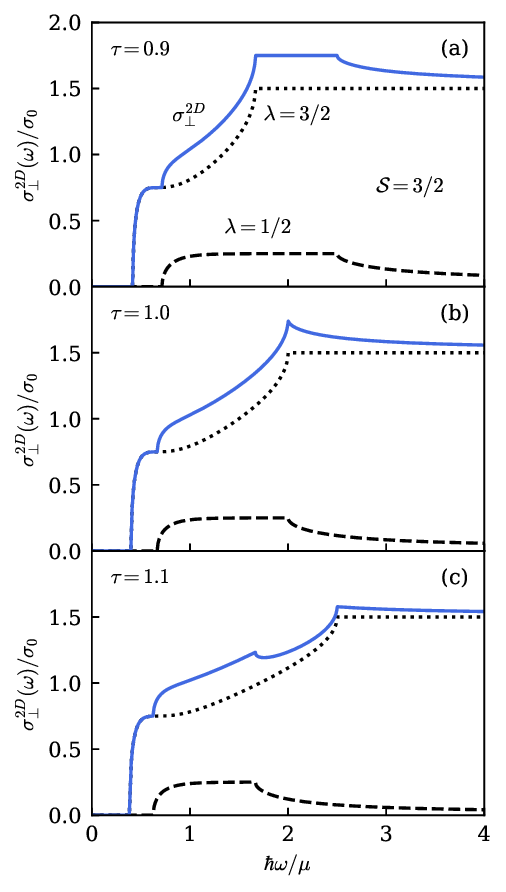}
    \caption{Optical conductivity perpendicular to the tilt direction in the 2D $\spin = 3/2$ case, where the total conductivity (solid blue curve) has been decomposed into the $\lambda = 1/2$ (dashed curve) and $\lambda = 3/2$ (dotted curve) contributions. To illustrate the interaction between the different energy scales or breakpoints, we present the cases of (a) $\tilt = 0.9$, (b) $\tilt = 1.0$, and (c) $\tilt = 1.1$. In all three plots, the $\lambda = 3/2$ cones are undertilted while the $\lambda = 1/2$ cones are overtilted, showing the varied responses that are possible in this intermediate case.}\label{fig:perp_decomposed}
\end{figure}

The behavior of overtilted cones (with open Fermi surfaces) can also be understood physically via this momentum alignment phenomenon. \cite{wildOpticalAbsorptionTwodimensional2022}
The $\spin = 3/2$ and $\spin=2$ cases discussed here are particularly interesting, since a scenario where one cone is overtilted and the others are undertilted (or where both are overtilted) can easily be constructed by tuning the tilt parameter $\tilt$.
These different scenarios can lead to a range of varied optical fingerprints as the two distinct responses mix. We refer to this scenario as type IV.
In Fig. \ref{fig:2D_32tilts}, we present optical conductivity results for 2D $\spin = 3/2$ materials with at least one overtilted cone, showing some of these possible optical fingerprints (for the undertilted case, see Fig. \ref{fig:slight_tilt2d}).
Notice that while the general shape of $\sigma_\parallel$ is relatively steady (monotonic increasing), $\sigma_\bot$ is much more varied and shows multiple instances of changing curvature and slope.
To provide intuition about this more complicated behavior, Fig. \ref{fig:perp_decomposed} presents only $\sigma_\bot$, decomposed into the contributions from the transitions ending at the $\lambda = 1/2$ and $\lambda = 3/2$ cones (black dashed and dotted lines, respectively).
In the overtilted $\lambda = 1/2$ case, both conductivity polarizations are zero up to $\hbar\omega = \frac{\mu}{\tilt+\lambda}$ just as in the undertilted case; however, $\mathcal{F}_\parallel$ is monotonic and asymptotically increases (see Fig. \ref{fig:2D_32tilts}), while $\mathcal{F}_\bot$ switches from increasing to asymptotically decreasing after the upper energy scale or breakpoint at $\hbar\omega=\frac{\mu}{\tilt-\lambda}$ (see Fig. \ref{fig:perp_decomposed}, dashed curves). 
The momentum alignment argument shows that for overtilted cones, the low energy breakpoint still corresponds to the first unblocked transitions (which are still `blue', and why the low energy behavior below $\hbar\omega=\frac{\mu}{\tilt-\lambda}$ appears similar to the undertilted case) but now the upper breakpoint corresponds to the photon energy at which transitions begin to be lost due to the valence band's open (hyperbolic) Fermi surface.
The states which are `lost' earliest are also `blue,' \textit{i.e.} contribute strongly to $\sigma_\bot$ and only weakly to $\sigma_\parallel$, producing the observed sharp drop in $\sigma_\bot$ while leaving $\sigma_\parallel$ relatively unaffected.
This more complicated behavior in the $\spin = 3/2$ case is therefore due to the interplay between the different energy breakpoints. An illustrative example is the $\tilt = 1.0$ case [Fig. \ref{fig:perp_decomposed}(b)], which shows a `cusp' in the combined conductivity where the two upper breakpoints are equal at $\hbar\omega = 2\mu$ [how these breakpoints interact for $\tilt$ different from 1 is shown in Figs. \ref{fig:perp_decomposed}(a) and \ref{fig:perp_decomposed}(c)].
The case where both cones are overtilted [Fig. \ref{fig:2D_32tilts}(c)] shows two `drops' corresponding to `losing' transitions in both cones.
Further discussion of the generic behavior of $\mathcal{F}_i(\lambda,\omega)$ in the under and overtilted cases is presented in Appendix \ref{app:generic}.

In experiment, these more varied behaviors may allow the sample's tilt orientation to be distinguished.
However, the pure and zero-temperature results presented here will be subject to some degree of broadening in experimental results, which may obscure some of the fine spectral details.
In Appendix~A, we give a brief discussion of how the $\spin>2$ cases, in particular, are likely to be affected by this.

\section{Tilted 3D Interband Results\label{sec:3D_inter}}
\label{sec:3Dtilted}


The interband optical conductivity of 3D materials with a set of untilted $\spin = 1/2$ Weyl cones is similar to that produced by the 2D case; the response is still zero up to a step at $\hbar\omega=2\mu$ (due to Pauli blocking) but above this step the spectrum is now linear instead of constant  due to the increased dimensionality reflected in the density of states \cite{timuskThreedimensionalDiracFermions2013}.
Tilted 3D $\spin = 1/2$ materials have been found to show a similar polarization-dependent splitting behavior to their 2D counterparts \cite{carbotteDiracConeTilt2016a,sonowalGiantOpticalActivity2019,houEffectsSpatialDimensionality2023}.
We have verified these $\spin = 1/2$ results and explicitly calculated the optical conductivity of the tilted $\spin = 1$, $3/2$, and 2 cases.
In this section, we present and discuss those results.

Similar to Eq.~(\ref{eq:2D_fullinter}), our calculations for a 3D material with pseudospin up to $\spin=2$ allows us to arrive at a general form for the optical conductivity which can be written as
\begin{equation}
    \sigma^{\textrm{3D}}_{i}(\omega) = A\omega \sum_{\lambda = \lambda_{min}}^\spin \lambda \mathcal{F}^{\textrm{3D}}_{i}(\lambda,\omega). \label{eq:3D_fullinter}
\end{equation}\newcommand{\internormThree}{\sigma_\omega \omega}
For an undertilted (type-I, $\tilt<\lambda$) cone, we have in the 3D case
\begin{subequations}
\begin{equation}
    \mathcal{F}_{\parallel}(\lambda,\omega) = \begin{cases}
        0 & \omega \leq  \minTI \\
        \frac{1}{4} \left(\xi_{-\lambda}^3-3\xi_{-\lambda} + 2\right) & \minTI < \omega \leq \maxTI \\
        1 & \maxTI < \omega, 
    \end{cases}
\end{equation}
\begin{equation}
    \mathcal{F}^{\textrm{3D}}_{\bot}(\lambda,\omega) = \begin{cases}
        0 & \omega \leq  \minTI \\
        \frac{1}{8} \left(-\xi_{-\lambda}^3-3\xi_{-\lambda} + 4\right) & \minTI < \omega \leq \maxTI \\
        1 & \maxTI < \omega.
    \end{cases}
\end{equation}\label{eq:3D_para_undertilted}
\end{subequations}
For overtilted (type-II, $\tilt>\lambda$) cones, we have
\begin{subequations}
\begin{equation}
    \mathcal{F}^{\textrm{3D}}_{\parallel}(\lambda,\omega) = \begin{cases}
        0 & \omega \leq  \minTII \\
        \frac{1}{4} \left(\xi_{-\lambda}^3-3\xi_{-\lambda} +2\right) & \minTII< \omega \leq \maxTII \\
        \frac{1}{4} (\xi_{-\lambda}^3-3\xi_{-\lambda}& \\
        \qquad - \xi_{+\lambda}^3+3\xi_{+\lambda}) & \maxTII < \omega,
    \end{cases}
\end{equation}
\begin{equation}
    \mathcal{F}^{\textrm{3D}}_{\bot}(\lambda,\omega) = \begin{cases}
        0 & \omega \leq  \minTII \\
        \frac{1}{8} \left(-\xi_{-\lambda}^3-3\xi_{-\lambda} + 4\right) & \minTII < \omega \leq \maxTII \\
        \frac{1}{8} (-\xi_{-\lambda}^3-3\xi_{-\lambda} & \\
        \qquad + \xi_{+\lambda}^3+3\xi_{+\lambda}) & \maxTII < \omega.
    \end{cases}
\end{equation}
\end{subequations}

\begin{figure}[ht]
    \includegraphics[width=\columnwidth]{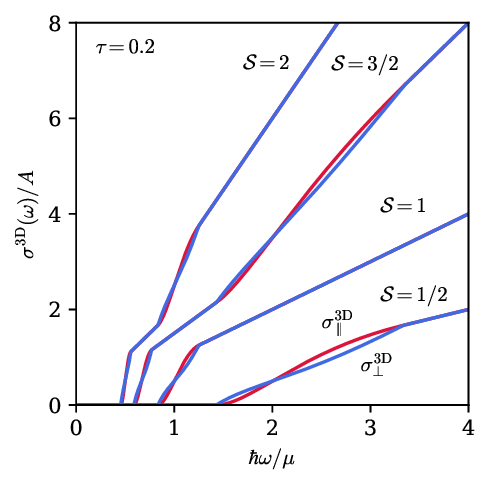}
    \caption{Frequency-dependent interband optical conductivity for slightly tilted (type-I) 3D Weyl materials with pseudospin $\spin = 1/2$, 1, $3/2$ and 2, where we have set $\tilt = 0.2$. Notice that we see a similar directional dependence of features as in the 2D case (Fig. \ref{fig:slight_tilt2d}) but that the directional dependence is less prominent due to the linear optical background.}\label{fig:slight_tilt3d}
\end{figure}

\begin{figure}[ht]
    \includegraphics[width=\columnwidth]{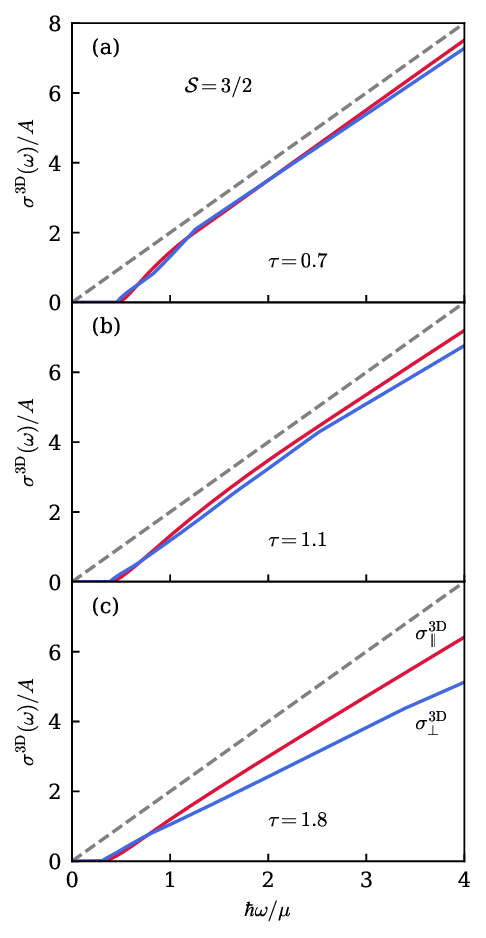}
    \caption{Interband optical conductivity for  type~IV overtilted 3D $\spin = 3/2$ Weyl materials, analogous to Fig. \ref{fig:2D_32tilts} [\textit{i.e.} (a,b) show results for one undertilted and one overtilted cone, while (c) shows the results when both cones are overtilted]. For comparison, the dashed lines plot saturated slope from all transitions in the undertilted case.
    }\label{fig:3D_32tilts}
\end{figure}

Finally, the critically tilted (type-III, $\tilt=\lambda$) case can still be viewed as the limiting case of the overtilted result, where the upper energy scale approaches infinity, yielding
\begin{subequations}
\begin{equation}
    \mathcal{F}^{\textrm{3D}}_{\parallel}(\lambda,\omega) = \begin{cases}
        0 & \omega \leq  \frac{\mu}{2\lambda} \\
        \frac{1}{4} \left(\xi_{-\lambda}^3-3\xi_{-\lambda} + 2\right) & \frac{\mu}{2\lambda} < \omega,
    \end{cases}
\end{equation}
\begin{equation}
    \mathcal{F}^{\textrm{3D}}_{\bot}(\lambda,\omega) = \begin{cases}
        0 & \omega \leq  \frac{\mu}{2\lambda} \\
        \frac{1}{8} \left(-\xi_{-\lambda}^3-3\xi_{-\lambda} + 4\right) & \frac{\mu}{2\lambda} < \omega. 
    \end{cases}
\end{equation}\label{eq:3D_perp_critical}
\end{subequations}
The same pattern as in the 2D case (Section~\ref{sec:2D_inter}) emerges in Eqs. (\ref{eq:3D_para_undertilted})-(\ref{eq:3D_perp_critical}).
Taking $\lambda = 1/2$, these forms reduce to those found in the literature for $\spin = 1/2$ \cite{carbotteDiracConeTilt2016a,houEffectsSpatialDimensionality2023,sonowalGiantOpticalActivity2019}.
We plot and discuss the features of $\mathcal{F}_\parallel$ and $\mathcal{F}_\bot$ for generic $\lambda$, with comparison to the 2D case, in Appendix \ref{app:generic} \cite{carbotteDiracConeTilt2016a,houEffectsSpatialDimensionality2023,sonowalGiantOpticalActivity2019}.

Fig. \ref{fig:slight_tilt3d} compares the interband conductivity results for 3D materials for the cases of $\spin = 1/2$, 1, $3/2$ and 2 cones.
Due to the linear background, the structure provided by the tilt is more difficult to discern, and the primary effect is to modify more directly the slope of the curves; $\sigma_\parallel$ and $\sigma_\bot$ are therefore less distinct. As a result of the tilt causing broadening to the vertical steps seen in the untilted case shown Fig.~\ref{fig:2D_3D_untilted}(b), the tilted curves will appear to have additional quasilinear line segments and the quasilinear sections of the curves arising from tilting will appear to extrapolate to negative intercepts rather than to the origin. These features may aid with the interpretation of such behaviors seen in some experiments. At this point, we wish to reiterate that our results are presented for $T=0$. With sufficient temperature the kink structure can be smoothed out as seen in Fig.~3 of Ref.~\cite{Singh:2021} for $T=0.4\mu$, however for $T=0.01\mu$ the kink structure is still seen. Experiments on graphene, done at 45K and with $\mu>100$ meV, give $T<0.04\mu$, indicating that it is possible for Dirac-Weyl systems to be in a regime where structure might be seen.

Just like in the 2D case, the optical conductivity can show type~IV `mixing' between different behaviors (undertilted, critically tilted, and overtilted) in the $\spin = 3/2$ and $\spin=2$ cases.
Fig. \ref{fig:3D_32tilts} demonstrates some of these potential fingerprints for $\spin=3/2$. It is clear that the multiple slopes seen in the undertilted case are gone and the net effect is to see something approaching a single quasilinear curve extrapolating to a negative intercept.
These results also show that $\sigma_\parallel$ and $\sigma_\bot$ are still similar in magnitude and slope in much of the plotted range [with the exception of the extremely overtilted $\tilt = 1.8$ case, Fig. \ref{fig:3D_32tilts}(c)].
As a result, it may prove difficult in experiment to confidently resolve the tilted and untilted directions using only interband optical conductivity measurements.

\section{Intraband Results\label{sec:intra}}

The low-frequency (Drude) part of the optical conductivity is due to intraband energy transitions at the Fermi level. In this section, we present this low frequency portion of the optical conductivity in both 2D and 3D (including a tilt), which was calculated alongside the interband results using Eq.~(\ref{eq:kubo_wavefunction}).
The net result is given in terms of a sum of Drude weights $W$ for each band which crosses the Fermi level, \textit{i.e.}
\begin{equation}
    \sigma_i(\omega) = \delta(\omega) \left(W_i (0)\delta_{\lfloor\spin\rfloor,\spin} + \sum_{\lambda = \lambda_{min}}^\spin W_i (\lambda)\right), \label{eq:full_intra}
\end{equation}
where the forms for the $W$'s are different for 2D and 3D (see below).
Here, the first term is contributed by the planar band (if it exists, \textit{i.e.} the $\spin$ must be integer) and the second term is a sum over the (tilted) conical bands.
In the case of undertilted cones, the Fermi surface of the conduction band is closed (and the valence band has no free states) so a closed form result can be directly obtained.
However, for overtilted (critically tilted) cones, each band contributes an open hyperbolic (parabolic) Fermi surface, with intraband transitions allowed for all states out to infinity; the planar band also contributes an infinite linear (2D) or planar (3D) Fermi surface for all values of $\tilt$.
A momentum cutoff must therefore be introduced to obtain closed form results in these cases.
A model yielding a more detailed structure of the complete band structure of a material beyond the low energy Dirac-Weyl approximation is therefore required to obtain a full picture of a given material's Drude response.

In the following, for clarity, we introduce a normalized $\taul = \tilt/\lambda$ and $\muell = \mu/\lambda$.
The closed-form 2D undertilted (type-I, $\taul < 1$) intraband results can be calculated without reference to a cutoff, and are given by
\begin{subequations}
\begin{align}
    W^{\textrm{2D}}_\parallel(\lambda) &= \frac{4\sigma_0 \lambda \muell}{\taul^2} \left(1 - \sqrt{1-\taul^2}\right),\label{eq:drude_2_under_para} \\
    W^{\textrm{2D}}_\bot(\lambda) &= \frac{4\sigma_0 \lambda \muell}{\taul^2} \frac{1 - \sqrt{1-\taul^2}}{\sqrt{1-\taul^2}}.\label{eq:drude_2_under_perp}
\end{align}
\end{subequations}
Notice that in the limit $\taul \rightarrow 0$, Eqs. (\ref{eq:drude_2_under_para}) and (\ref{eq:drude_2_under_perp}) yield $W^{\textrm{2D}}_{(\parallel,\bot)}(\lambda) = 2\mu\sigma_0$ irrespective of any specific $\lambda$, as expected.
In the cases of the tilted planar band and the cones which are critically tilted or overtilted, the introduction of a cutoff $\Lambda$ is required to obtain closed form results. 
For overtilted cones, we follow Tan et al. \cite{tanSignaturesLifshitzTransition2022} and define
\begin{align}
    A(\mu,\tilt,\Lambda, \lambda) &= \frac{\Lambda}{\muell \taul} \sum_{\chi = \pm} \sqrt{1-\left(\frac{\muell-\chi\Lambda}{\taul\Lambda}\right)^2}, \\
    \begin{split}
    B(\mu,\tilt,\Lambda, \lambda) &= \frac{1}{\taul^2} \sum_{\chi = \pm} \chi \ln \left[\taul^2 + \chi \frac{\muell-\chi \Lambda}{\Lambda} \right. , \\
    &+ \left.\sqrt{\taul^2-1}\sqrt{\taul^2 - \left(\frac{\muell-\chi \Lambda}{\Lambda}\right)^2}\right]
    \end{split} \\
    C(\mu,\tilt,\Lambda, \lambda) &= \frac{1}{\taul^2} \sum_{\chi = \pm} \arccos\left(\frac{\muell-\chi\Lambda}{\taul\Lambda}\right).
\end{align}
The results for the type-II, $\taul>1$ case are therefore
\begin{subequations}
\begin{align}
    \begin{split}
    W^{\textrm{2D}}_\parallel(\lambda) &= 4\sigma_0 \lambda \muell \left[ (\taul^2-1)A(\mu,\tilt,\Lambda, \lambda)\right. + \\
    &\left.\sqrt{\taul^2-1}B(\mu,\tilt,\Lambda, \lambda) + C(\mu,\tilt,\Lambda, \lambda)\right],
    \end{split} 
    \\
    \begin{split}
    W^{\textrm{2D}}_\bot(\lambda) &= 4\sigma_0 \lambda \muell \left[ A(\mu,\tilt,\Lambda, \lambda)\right. \\
    &+\left.\frac{1}{\sqrt{\taul^2-1}}B(\mu,\tilt,\Lambda, \lambda) - C(\mu,\tilt,\Lambda, \lambda) \right].
    \end{split}
\end{align}
\end{subequations}
For critically tilted (type-III, $\taul = 1$) cones, we have
\begin{subequations}
\begin{align}
    W^{\textrm{2D}}_\parallel(\lambda) &= 4\sigma_0 \lambda \muell \frac{1}{\pi}\arccos\left(\frac{\muell-\Lambda}{\Lambda}\right), \\
    W^{\textrm{2D}}_\bot(\lambda) &= 4\sigma_0 \lambda \muell \frac{2}{\pi}\left[ \sqrt{\frac{2\Lambda-\muell}{\muell}}-\arccos\sqrt{\frac{\muell}{2\Lambda}} \right].
\end{align}
\end{subequations}
Finally, the planar band contributes
\begin{subequations}
\begin{align}
    W^{\textrm{2D}}_\parallel(0) &= 4\sigma_0 \mu \frac{1}{\pi}\sqrt{\frac{\Lambda^2 \tau^2}{\mu^2} - 1}~\Theta(\Lambda - \mu/\tau), \\
    W^{\textrm{2D}}_\bot(0) &= 0,
\end{align}
\end{subequations}
for all values of $\tilt$, where the Heaviside function ensures that the planar band only contributes if the linear Fermi surface is within the cutoff.

\newcommand{\zintranorm}{\frac{3}{2}A}
\newcommand{\xintranorm}{\frac{3}{4}A}
The 3D undertilted Drude weights are
\begin{subequations}
\begin{align}
    W^{\textrm{3D}}_\parallel(\lambda) &= \zintranorm \lambda \frac{\muell^2}{\taul^3} \left[\ln\left(\frac{1+\taul}{1-\taul}\right)-2\taul\right], \\
    W^{\textrm{3D}}_\bot(\lambda) &= \xintranorm \lambda \frac{\muell^2}{\taul^3}\left[\ln\left(\frac{1-\taul}{1+\taul}\right)+\frac{2\taul}{1-\taul^2}\right].
\end{align}
\end{subequations}
Unlike the 2D case, these 3D undertilted results are $\lambda$-dependent in the limit $\taul \to 0$, yielding $W^{\textrm{3D}}_{(\parallel, \bot)}(\lambda) = \mu^2 A /\lambda$. 
The overtilted results are
\begin{subequations}
\begin{align}
    \begin{split}
    &W^{\textrm{3D}}_\parallel(\lambda) = \zintranorm \lambda \frac{\muell^2}{\taul^3}\left[ \ln\left(\frac{\Lambda^2}{\muell^2}\right)\right. \\
    &+\left. \ln\left(\taul^2-1\right) + \frac{\Lambda^2}{\muell^2}\left(\taul^2-1\right)^2 + 3 - \taul^2 \right],
    \end{split}
    \\
    \begin{split}
    &W^{\textrm{3D}}_\bot(\lambda) = \xintranorm \lambda \frac{\muell^2}{\taul^3}\left[ \ln\left(\frac{\muell^2}{\Lambda^2}\right)\right. \\
    &- \left.\ln\left(\taul^2-1\right) + \frac{\Lambda^2}{\muell^2}\left(\taul^2-1\right) - \frac{\taul^2-3}{\taul^2-1} \right].\label{eq:drude_3dperp_over}
    \end{split}
\end{align}
\end{subequations}
The critically tilted case yields
\begin{subequations}
\begin{align}
    W^{\textrm{3D}}_\parallel(\lambda) &= \zintranorm \lambda \muell^2 \left[ \ln\left(\frac{2\Lambda}{\muell}\right) \right], \\
    W^{\textrm{3D}}_\bot(\lambda) &= \xintranorm \lambda \muell^2 \left[ \frac{2\Lambda}{\muell} - \ln\left(\frac{2\Lambda}{\muell}\right) - 1 \right].\label{eq:drude_3dperp_crit}
\end{align}
\end{subequations}
Finally, the hyperplanar band yields
\begin{subequations}
\begin{align}
    W^{\textrm{3D}}_\parallel(0) &= \frac{3}{4}A\frac{\mu^2}{\tau}\left[\frac{\Lambda^2 \tau^2}{\mu^2} - 1\right] \Theta(\Lambda - \mu/\tau), \\
    W^{\textrm{3D}}_\bot(0) &= 0,
\end{align}
\end{subequations}
for all values of $\tilt$.
Note that the flat band for both dimensionalities only contributes to $W_\parallel$.

For the case of tilted $\spin=1/2$, we recover similar results to Ref.~\cite{houEffectsSpatialDimensionality2023}
Using the results presented here and those derived for the interband conductivity in the previous sections, we can now discuss the implications, in the undertilted case, for optical sum rules in the following section.

\section{Velocity Anisotropy and Sum Rules\label{sec:sum_rules}}


The results presented so far in this work have been for materials with isotropic cones with an added tilting term, \textit{i.e.} where the Fermi velocity in the untilted Hamiltonian is the same in all directions; however, both 2D and 3D materials may also have anisotropic tilted cones.
Additionally, the optical conductivity results presented in previous sections can follow sum rules for spectral weight transfer and other interesting mathematical relations which may allow for the extraction of information from the optical data.
In this section, we discuss these issues.
When discussing sum rules, we focus on only untilted and undertilted cones, as the open Fermi surface in the overtilted case prevents a coherent analysis; the basic effect of anisotropy, however, holds for all tilt phases.

To model anisotropic cones, we consider a modification to Eq.~(\ref{eq:basic_ham}) of the form
\begin{equation}
    \ham = \hbar (v_x k_x \ophat{S}_x + v_y k_y \ophat{S}_y + v_z k_z \ophat{S}_z + v_\tilt k_\parallel \ophat{S}_0), \label{eq:aniso_ham}
\end{equation}
where we have use the notation of tilt velocity $v_\tilt$ in place of $v\tau$ of Eq.~(\ref{eq:basic_ham}). The $k_i$ and $k_\parallel$ are as defined previously. From this we calculate the conductivity, which we will label by $\tilde\sigma_{ii}(\omega)$. For $v_x=v_y=v_z=v$, we recover Eq.~(\ref{eq:basic_ham}) and our previous conductivity results $\sigma_{ii}(\omega)$. 
In the case of velocity anisotropy, previous authors (e.g., \cite{wildOpticalAbsorptionTwodimensional2022,houEffectsSpatialDimensionality2023}) have shown that in the 2D spin-1/2 case,
\begin{equation}
    \tilde\sigma_{ii}(\omega) = \frac{v_{i}^2}{v_x v_y}\sigma_{ii}(\omega),
\end{equation}
and for 3D spin-1/2 case,
\begin{equation}
    \tilde\sigma_{ii}(\omega) = \frac{v_{i}^2v}{v_x v_y v_z}\sigma_{ii}(\omega).
\end{equation}
We have verified that the same result extends to higher pseudospin.
In the untilted and undertilted cases, the qualitative effect of this anisotropy is to provide a directional dependence to the saturated optical background at large $\hbar\omega$ beyond $\hbar\omega=2\mu$ (for half-integer pseudospin) or $\hbar\omega=\mu$ (for integer pseudospin).
These differing backgrounds may make experimental analysis more subtle, but these simple changes from the isotropic behavior at large $\omega$ mean that relatively simple rules may allow the data to be re-mapped back to an isotropic case.

Graphene is known to have an optical sum rule; any `missing' spectral weight in the interband portion due to a finite chemical potential $\mu$ is transferred to the intraband weight. This is a statement that
the integral over the entire conductivity yields the same value independent of $\mu$ \cite{gusyninMagnetoopticalConductivityGraphene2006}.
This sum rule has also been found to be valid in the undertilted spin-1/2 case for both the 2D and 3D cases \cite{carbotteDiracConeTilt2016a}.
For the $\spin = 3/2$ case, we can immediately confirm that this sum rule holds in both polarization directions.
Indeed, we can write the sum rule for each undertilted cone ($\lambda>0$) as:
\begin{equation}
    W_i^{\textrm{2D}}(\lambda) = 2 \lambda \sigma_0 \int_0^{\mu/(\lambda-\tau)} \left[1-\mathcal{F}^{\textrm{2D}}_i(\lambda,\omega)\right]d\omega,
\end{equation}
\begin{equation}
    W_i^{\textrm{3D}}(\lambda) = 2 \lambda A \int_0^{\mu/(\lambda-\tau)} \omega \left[1-\mathcal{F}^{\textrm{3D}}_i(\lambda,\omega)\right]d\omega.
\end{equation}
For $\spin = 1$ and 2, however, the Drude contribution from the cones completely accounts for the missing interband spectral weight; the planar band's excess contribution in the $k_\parallel$ direction therefore breaks this sum rule, while it is preserved in $k_\bot$ directions [since $W_\parallel(\lambda = 0)\neq 0$ while $W_\bot(\lambda = 0) = 0$]. The Fermi surface provided by the tilted planar band is now an open one (a line in 2D, for example).
A broken sum rule along one direction but not in perpendicular directions is a characteristic of integer $\spin$.
Generally, these results are anisotropy-independent, since a multiplicative factor is simply applied to parts of the conductivity in each direction.

For undertilted cones, the ratio of different directions' velocities $v_i/v_j$ can be extracted from the high frequency optical conductivity (where the constant or linear background is saturated), since
\begin{equation}
    \sqrt{\frac{\tilde\sigma_{ii}(\omega_s)}{\tilde\sigma_{jj}(\omega_s)}} = \frac{v_{i}}{v_{j}},
\end{equation}
where $\omega_s$ represents a frequency at which the response is saturated (constant or linear for 2D or 3D, respectively).
In the 2D case, a corollary is that 
\begin{equation}
    \sqrt{\tilde\sigma_{ii}(\omega_s)\tilde\sigma_{jj}(\omega_s)} = \sigma(\omega_s), \label{eq:multiply_rule}
\end{equation}
\textit{i.e.} the saturated isotropic conductivity can be extracted from the two anisotropic directions (in the 3D case, the right hand side of Eq.~(\ref{eq:multiply_rule}) would be altered to $\sigma(\omega_s)v/v_\ell$ where $\ell\neq i$ or $j$).
Additionally, another combination that we have found which maybe useful for experimental analysis is one that uses the tilt-broadened region. Defining
\begin{equation}
  I_{ii}(m)=  \int_{\omega_{min}}^{\omega_{max}}  \frac{\tilde\sigma_{ii}^{\rm inter}(\omega)}{\omega^m}d\omega,
\end{equation}
then
\begin{equation}
  \sqrt{\frac{I_{\bot}(m)}{I_{\parallel}(m)}}
    = \frac{v_{\bot}}{v_{\parallel}},
\end{equation}
where $m=2$ or 3, for 2D or 3D, respectively. 
Here, $\omega_{min}$, and $\omega_{max}$ are the lowest and highest energy breakpoints in the optical conductivity spectrum. This is based on the result that for 2D
\begin{equation}
  \int_{\omega_{min}}^{\omega_{max}}  \frac{\mathcal{F}^{\textrm 2D}_i(\lambda,\omega)}{\omega^2}d\omega=\frac{\hbar\tau}{\mu},
\end{equation}
and for 3D
\begin{equation}
  \int_{\omega_{min}}^{\omega_{max}}  \frac{\omega\mathcal{F}^{\textrm 3D}_i(\lambda,\omega)}{\omega^3}d\omega=\frac{\hbar\tau}{\mu},
\end{equation}
recalling in this latter case that the conductivity has an extra factor of $\omega$ in front of the sum over $\lambda$ in Eq.~(\ref{eq:3D_fullinter}) that needs to enter the integral, which we make explicit here. Moreover, these results also imply that for 2D, $I_{ii}(2)/\tilde\sigma_{ii}(\omega_s) = \hbar\tau/\mu$ and likewise for 3D, $\omega_s I_{ii}(3)/\tilde\sigma_{ii}(\omega_s) = \hbar\tau/\mu$, which could be another way of extracting the information on the tilt parameter using the weighted integration of the broadened region in ratio with the higher frequency saturated background value.

\section{Tilted $\alpha$-{\cal T}$_3$ Model\label{sec:a-T3}}

\begin{figure}[ht]
    \includegraphics[width=\columnwidth]{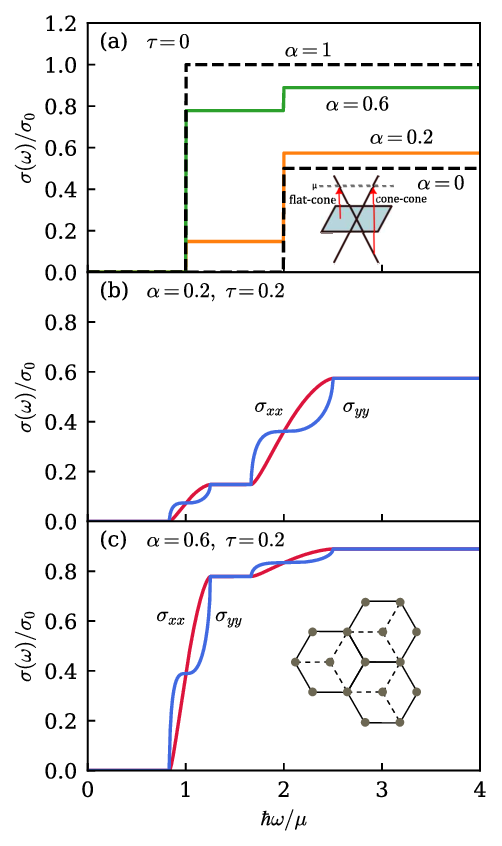}
    \caption{Interband optical conductivity of undertilted $\alpha$-{\cal T}$_3$ model. (a) $\tau = 0$, showing the untilted energy scales and backgrounds for different values of $\alpha$. The inset is a diagram of the interband transitions, which are now allowed from cone to cone or from the flat band to a cone (Ref. \cite{illesHallQuantizationOptical2015}). Fixing $\tilt = 0.2$ (undertilted), we show varying $\alpha$: (b) $\alpha = 0.2$ and (c) $\alpha = 0.6$. Inset in (c) is the $\alpha$-{\cal T}$_3$ lattice, with solid lines representing linkages with a hopping parameter $t_h$ and dashed lines a hopping parameter $\alpha t_h$.\label{fig:a-T3_under}}
\end{figure}

As discussed previously, a nearest-neighbor tight-binding hopping model applied to the honeycomb lattice results in the $\spin=1/2$ model at low energy. Likewise,  the $\spin=1$ case can arise from a similar analysis on a dice (or {\cal T}$_3$) lattice. The $\alpha$-{\cal T}$_3$ model\cite{Piechon:2015} interpolates between the honeycomb and dice lattices by introducing a variable hopping parameter $\alpha t_h$ ($0\leq\alpha\leq1$) linking one set of sites in the dice lattice, whereas the other linkages have a hopping parameter $t_h$  (see inset of Fig. \ref{fig:a-T3_under}(c)).
Here, we add to previous literature on this model \cite{illesHallQuantizationOptical2015,illesMagneticPropertiesEnsuremath2016, carbotteOpticalPropertiesSemiDirac2019} by calculating the optical conductivity arising from the $\alpha$-{\cal T}$_3$ Hamiltonian with the inclusion of a tilting term:
\begin{widetext}
\begin{equation}
    \ham_{\alpha\text{-T}_3} = \hbar v_F k \left(\begin{matrix}
        0 & e^{-i\pp}\cos\theta & 0 \\
        e^{i\pp}\cos\theta & 0 & e^{-i\pp}\sin\theta \\
        0 & e^{i\pp}\sin\theta & 0
    \end{matrix}\right) + \hbar v_F k\cos{\pp} \tilt \opS_0, \label{eq:ham_aT3}
\end{equation}
\end{widetext}
where $\alpha = \tan\theta$ and $\cos\pp = k_x/k$.
The first term in Eq.~(\ref{eq:ham_aT3}) is the untilted $\alpha$-{\cal T}$_3$ Hamiltonian, while the second term applies the tilt in the $x$-direction. (Note, the notational use of $\theta$ and $\phi$ is reversed to that of Ref.~\cite{illesHallQuantizationOptical2015}.)
The resulting eigenvalues are those of the tilted 2D, $\spin=1$ case,
\begin{equation}
    \epsilon_n(\vec{k}) = \hbar v_F k(\tilt\cos\pp + n),
\end{equation}
where $n = 0,\pm1$, while the wavefunctions are unchanged from the untilted $\alpha$-{\cal T}$_3$ case (cf. Ref. \cite{illesHallQuantizationOptical2015}),
\begin{equation}
    \ket{0} = \left(\begin{matrix}
        \sin\theta e^{i\pp} \\
        0 \\
        -\cos\theta e^{-\pp}
    \end{matrix}\right), ~~~ \ket{\pm} = \frac{1}{\sqrt{2}}\left(\begin{matrix}
        \cos\theta e^{i\pp} \\
        \pm1 \\
        \sin\theta e^{-i\pp}
    \end{matrix}\right).
\end{equation}
The relevant matrix elements for our calculation of the optical conductivity are
\begin{equation}
    \abs*{\mel{0}{\ophat{v}_x}{0}}^2 = \tilt^2,
\end{equation}\begin{equation}
    \abs*{\mel{\pm}{\ophat{v}_x}{\pm}}^2 = (\tilt \pm \cos\pp)^2,
\end{equation}\begin{equation}
    \abs*{\mel{0}{\ophat{v}_x}{\pm}}^2 = \frac{1}{2}\sin^2(2\theta)\sin^2\pp,
\end{equation}\begin{equation}
    \abs*{\mel{\mp}{\ophat{v}_x}{\pm}}^2 = \cos^2(2\theta)\sin^2\pp,
\end{equation}
and
\begin{equation}
    \abs*{\mel{0}{\ophat{v}_y}{0}}^2 = 0,
\end{equation}\begin{equation}
    \abs*{\mel{\pm}{\ophat{v}_y}{\pm}}^2 = \sin^2\pp,
\end{equation}\begin{equation}
    \abs*{\mel{0}{\ophat{v}_y}{\pm}}^2 = \frac{1}{2}\sin^2(2\theta)\cos^2\pp,
\end{equation}\begin{equation}
    \abs*{\mel{\mp}{\ophat{v}_y}{\pm}}^2 = \cos^2(2\theta)\cos^2\pp.
\end{equation}
In this model, in spite of the band structure appearing the same as the $\spin=1$ case, transitions can now occur between the cones, in addition to those between the flat band and a cone (see inset schematic in Fig.~\ref{fig:a-T3_under}(a))
Notice that elements associated with a transition between the the two cones have a factor of $\cos^2(2\theta)$ associated with them, while those associated with a transition between the flat band and the cones come with a factor of $\sin^2(2\theta)$; these factors are responsible for modulating the cone-to-cone and flat-to-cone portions of the conductivity, respectively.
The intraband elements are $\theta$-independent, a characteristic of the $\alpha$-{\cal T}$_3$ model.

\begin{figure}[ht]
    \includegraphics[width=\columnwidth]{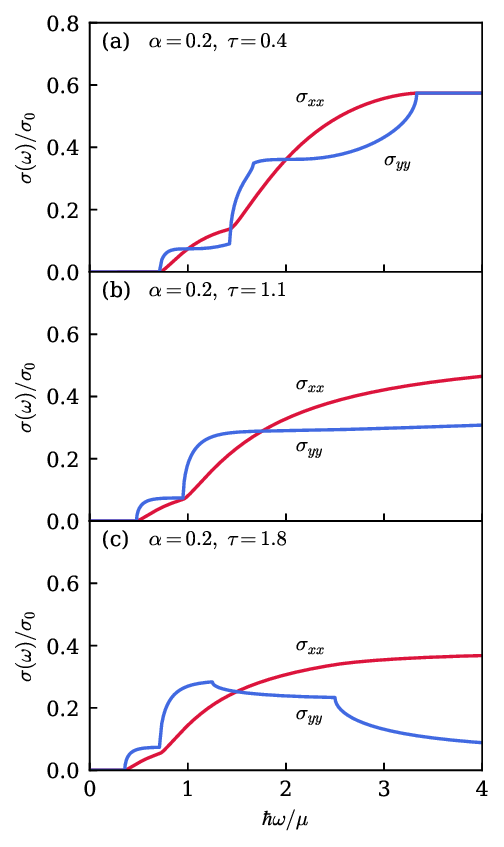}
    \caption{Interband optical conductivity of $\alpha$-{\cal T}$_3$ model comparing under and overtilted signatures, for fixed $\alpha = 0.2$ and varying $\tilt$: (a) $\tilt = 0.4$, (b) $\tilt = 1.1$, and (c) $\tilt = 1.8$.\label{fig:a-T3_3stack}}
\end{figure}

We proceed to calculate the optical conductivity using Eq.~(\ref{eq:kubo_wavefunction}). 
The resulting interband conductivities are
\begin{widetext}
\begin{subequations}
\begin{align}
    \sigma^{\alpha\text{-\cal T}_3}_{xx}(\omega) &= \sigma_0\left(\frac{1}{2}\cos^2(2\theta)\mathcal{F}^{\textrm{2D}}_\parallel(1,\omega,2\mu) + \sin^2(2\theta)\mathcal{F}^{\textrm{2D}}_\parallel(1,\omega,\mu)\right),\label{eq:inter_xx_aT3}\\
    \sigma^{\alpha\text{-\cal T}_3}_{yy}(\omega) &= \sigma_0\left(\frac{1}{2}\cos^2(2\theta)\mathcal{F}^{\textrm{2D}}_\bot(1,\omega,2\mu) + \sin^2(2\theta)\mathcal{F}^{\textrm{2D}}_\bot(1,\omega,\mu)\right).\label{eq:inter_yy_aT3}
\end{align}
\end{subequations}
\end{widetext}
The first and second terms in Eqs. (\ref{eq:inter_xx_aT3}) and (\ref{eq:inter_yy_aT3}) are the cone-to-cone and flat-to-cone type contributions, respectively.
We have used $\mathcal{F}^{\textrm{2D}}_i(1,\omega,2\mu)$
to indicate that the form for $\mathcal{F}^{\textrm{2D}}_i(\lambda,\omega)$ from Section \ref{sec:2D_inter} should be used here but with all occurrences of $\mu$ replaced by $2\mu$. This reflects that the band structure is still the same as the tilted $\spin =1$ case and hence $\lambda=1$, but that transitions occur between the valence and conduction cones and start at $2\mu$ (see inset of Fig.~\ref{fig:a-T3_under}(a)). 
The respective Drude weights are $\alpha$-independent, and are
\begin{subequations}
\begin{align}
    W^{\alpha\text{-\cal T}_3}_{xx} & = W^{\textrm{2D}}_\parallel(1) + W^{\textrm{2D}}_\parallel(0), \label{eq:drude_xx_aT3}\\
    W^{\alpha\text{-\cal T}_3}_{yy} & = W^{\textrm{2D}}_\bot(1), \label{eq:drude_yy_aT3}
\end{align}
\end{subequations}
\textit{i.e.} they are the same as the $\spin = 1$ case (as expected, since the matrix elements are independent of $\alpha$ and the band structure is the same).

Fig. \ref{fig:a-T3_under} plots the interband part of the conductivity in the undertilted case [Eqs. (\ref{eq:inter_xx_aT3}) and (\ref{eq:inter_yy_aT3})].
In Fig. \ref{fig:a-T3_under}(a), we seek to orient the reader with respect to the $\alpha$-dependence of the step heights and optical backgrounds by plotting the untilted case ($\tau = 0$) with various values of $\alpha$.
Notice that $\alpha = 0$ and $\alpha = 1$ reproduce the $\spin = 1/2$ and $\spin = 1$ interband results, respectively, while for intermediate values of $\alpha$ the optical conductivity interpolates between these two cases using the Berry's phase-dependent prefactors $\cos^2(2\theta)/2$ and $\sin^2(2\theta)$, where the Berry's phase $\Phi^{\rm B}=\pi\cos(2\theta)$ \cite{illesHallQuantizationOptical2015,carbotteOpticalPropertiesSemiDirac2019}.
Figs. \ref{fig:a-T3_under}(b) and \ref{fig:a-T3_under}(c) show a similar comparison for the $\tau = 0.2$ case, where we have taken (b) $\alpha = 0.2$ and (c) $\alpha = 0.6$ (a previous mapping between massless Kane fermions and the $\alpha$-{\cal T}$_3$ model using $\alpha = 1/\sqrt{3} \approx 0.6$ motivates this latter choice \cite{malcolmMagnetoopticsMasslessKane2015}).
The flat step heights here are the same as in the untilted cases, with the characteristic splitting of energy scales and directional dependence that we expect for tilted cones. Notice, however, that the broadening caused by the tilted energy scales about $\hbar \omega = 2\mu$ in Fig.~\ref{fig:a-T3_under}(b) is not as wide as seen in the $\spin=1/2$ case plotted in Fig.~\ref{fig:slight_tilt2d}.
This broadening in the $\alpha$-T$_3$ model is set by $\lambda=1$ and hence is less broad than would be for $\lambda=1/2$.
This makes clear that the tilted $\alpha$-T$_3$ model is not truly a weighted sum of a $\spin=1/2$ and $\spin=1$, but is based on the $\spin=1$ case with additional allowed cone-to-cone transitions.

The optical `fingerprints' produced here are similar to the $\spin = 3/2$ or $\spin = 2$ cases (see Appendix \ref{app:2D_higher} for a discussion of pseudospin beyond $\spin = 2$) in that the spectrum is composed of exactly two `steps' in the untilted conductivity spectrum.
However, the location of these steps and the heights of each step will distinguish $\alpha$-{\cal T}$_3$-type fingerprints from both the $\spin = 3/2$ and $\spin = 2$ cases.
In the tilted $\alpha$-{\cal T}$_3$ case, the breakpoints are centered around $\hbar \omega = \mu$ and $\hbar \omega = 2\mu$, and the relative flat step heights are set by the value of $\alpha$.
The breakpoints for each `pure' pseudospin are also centered around particular values, but the relative step heights are uniquely determined by the particular value of $\spin$.
In the $\spin = 3/2$ case, the two flat steps have a relative height of 3:1 (First:Second), and the breakpoints are centered around $\hbar \omega = 2\mu/3$ and $\hbar \omega = 2\mu$.
The $\spin = 2$ case has equal-height (1:1) steps, with breakpoints centered around $\hbar \omega = \mu/2$ and $\hbar \omega = \mu$.
Optical conductivity measurements not fitting these particular results cannot be attributed to one of these single $\spin$ value cases, however, the tilted $\alpha$-{\cal T}$_3$ model may provide an alternate explanation.
Most distinctively, the first flat step for the $\alpha$-{\cal T}$_3$ case may be \textit{smaller} than the second, \textit{e.g.} for $\tau = 0.2$ [Fig. \ref{fig:a-T3_under}(b)].
Notice that the ratio between step locations in $\hbar\omega$ based on $\tilt=0$ for $\spin = 2$ is the same as for $\alpha$-{\cal T}$_3$ (\textit{i.e.} they are a factor of 2 apart), and hence the ratio of step heights must be relied upon  as a distinguishing attribute if the chemical potential $\mu$ for the system is unknown.
The approximate values of $\alpha$ needed to give the same step height ratios as for $\spin = 3/2$ and $\spin = 2$ are $\alpha\approx 0.47$ and 0.32, respectively.

To emphasize the distinction between the $\alpha$-{\cal T}$_3$ and $\spin = 3/2$ cases, Fig. \ref{fig:a-T3_3stack} compares the $\tilt$-dependence of the $\alpha = 0.2$ case.
Note that the $\tilt = 0.4$ case [Fig. \ref{fig:a-T3_3stack}(a)] is still undertilted, but does not show two distinct flat `steps' in the conductivity spectrum like the $\tilt = 0.2$ case [see Fig. \ref{fig:a-T3_under}(b)] due to the interplay between the breakpoints.
The overtilted cases [Figs. \ref{fig:a-T3_3stack}(b) and \ref{fig:a-T3_3stack}(c)] also show similar signatures to the $\spin = 3/2$ cases where both cones are overtilted - though similar rules to the undertilted cases regarding absorption heights and breakpoint locations apply here.
A particularly interesting feature of these plots is that the weaker first step has introduced a kink at low frequency in the $\sigma_{xx}$ (\textit{i.e.} $\sigma_\parallel$, red) curves.
This kink is characteristic of the activation of the second transition type, and is `smoothed out' in cases where the first step is higher (e.g. the `pure' pseudospin cases seen earlier) due to the resolution of the plots.

Additionally, the spectral weight transfer sum rule still holds for the conical band part of the Drude weight at any undertilted value of $\tau$ (at a given value of $\alpha$ - different values of $\alpha$ have different saturated backgrounds and therefore different spectral weight transfer).
The net result is therefore a violation of the sum rule in the tilted direction but not in the untilted direction, just as in the $\spin = 1$ case.

\section{Summary\label{sec:summary}}

In conclusion, we have calculated the absorptive, frequency-dependent longitudinal conductivity for Dirac-Weyl Hamiltonians representing different pseudospin $\spin$ values, where a tilting term is included. The additional term has the effect of tilting the band structure. We have reviewed the case of $\spin=1/2$ for both 2D and 3D to set the stage for the explicit calculations and discussion that we provide for $\spin=1$, $3/2$, and 2, in order to demonstrate the signatures of tilting for higher pseudospin. Undertilted (type I), overtilted (type II), and critically tilted (type III) cases have been presented. In addition, we define a fourth scenario, which we call type IV, unique to higher $\spin$ having multiple nested cones, such as $\spin=3/2$ and $\spin=2$, where one cone may be identified as undertilted while a second cone is overtilted, thus showing an admixture of type I and type II fingerprints. This case, illustrates the rich behavior that can occur in the optical response. We also note that the cases of $\spin=1$ and $\spin=1/2$ can show similar qualitative behavior, but the energy scales will be different as well as the high-frequency saturated background. Likewise, a similar statement holds for comparing $\spin=3/2$ and $\spin=2$, etc. If knowledge of the chemical potential or the overall absolute value of the saturated conductivity is not known, then it may be difficult to provide an exact determination in experiment without further input from theory or other measurements. In the case of 3D, the extra linear background which modifies the conductivity tends to obscure the more subtle details of the effects of tilting and the resulting behavior is most likely to be seen as a series of quasilinear- or linear-in-$\omega$ responses, where at low energy the slope of the curve has a negative intercept. This is important as negative intercepts have been seen in experiments on possible Dirac-like materials and this effect cannot be produced in a simple Dirac model with no tilting. The number of linear sections seen in the conductivity curve will depend on the pseudospin value, but for $\spin=1$ and $\spin=1/2$, only the lowest segment has a negative intercept and the higher segment extrapolates to the origin. For $\spin=3/2$ and above, other segments of the curve will display a negative intercept. This is to be contrasted to the case without tilting in the band structure, where for any $\spin$ value, all linear segments will extrapolate to the origin. Our results are for the pure limit, but impurity scattering will only produce a shift upwards towards a positive intercept, never a negative one.

As we have provided the interband and intraband conductivity for higher pseudospin cases discussed here, we also examined the possibility of sum rules or spectral weight transfer for finite $\mu$. We have found that the standard optical sum rule for transfer of spectral weight, with doping $\mu$, continues to hold for undertilted $\spin=1/2$ and higher half-integer $\spin$ values in both the parallel and perpendicular directions. In the case of integer $\spin$, the sum rule will hold in the perpendicular direction but will not hold in the parallel direction. This violation occurs due to the tilted flat band providing an open Fermi surface which, like the overtilted cases, cannot obey the sum rule.

Finally, we examined the $\alpha$-T$_3$ model which is a hybrid of $\spin=1/2$ and 1 behaviors, resulting from a modification of an intersite hopping term in the dice lattice with a parameter $\alpha$. This  model has a variable Berry's phase and has found some possible support in experiment. In our results, we have found that the $\alpha$-T$_3$ optical response can have some similarities with both the $\spin=2$ and $\spin=3/2$ cases, but there are important differences in the details. Once again, if knowledge of the value of the chemical potential and the overall saturated background is lacking, it may be difficult to differentiate between $\spin=2$ and the $\alpha$-T$_3$ model for the case of $\alpha\sim 0.3$. For the case of $\spin=3/2$, ratios of important energy scales will allow for a distinction between the $\alpha$-T$_3$ model and $\spin=3/2$.

In summary, while materials with different pseudospin-$\spin$ character are mainly discussed within the realm of theory, it is hoped that providing the expected optical fingerprints of these models will aid in the eventual characterization of new materials. Optical response has been one of the important successful probes of graphene and excellent agreement between theory and experiment has been demonstrated. 

\begin{acknowledgments}
We thank Elijah Kent for discussions and thank B. Dor\'a answering questions about his work.
This work has been supported by the Natural Sciences and Engineering Research Council of Canada (Grant No. RGPIN-2017-03931).
\end{acknowledgments}

\appendix

\section{Higher Pseudospin Beyond $\spin=2$\label{app:2D_higher}}

\begin{figure}[ht]
    \includegraphics[width=\columnwidth]{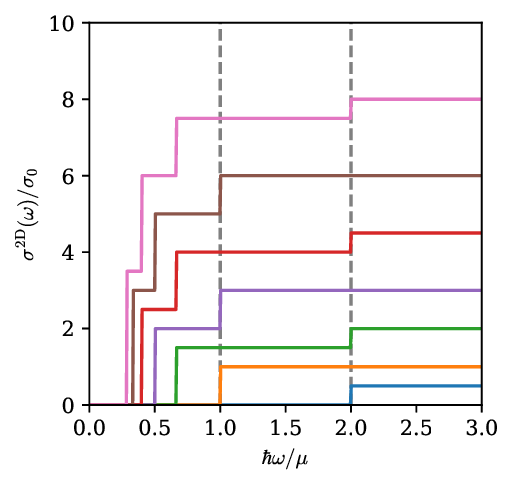}
    \caption{Untilted 2D interband optical conductivity for pseudospin up to $\spin = 7/2$, depicting the pattern of absorption steps beyond $\spin = 2$. The top curve (magenta) is for $\spin = 7/2$, followed in descending order by $\spin=3$ (brown), 5/2 (red), 2 (violet), 3/2 (green), 1 (orange), and finally $\spin= 1/2$ (blue) shown in the lowest curve. As $\spin$ increases, the high energy saturated conductivity increases and new steps appear at each new allowed value for $\hbar \omega = \mu/\lambda$. Note that the height of the lowest energy step is $\spin\sigma_0$ \cite{doraLatticeGeneralizationDirac2011a}. The vertical grey dashed lines indicate $\hbar\omega=\mu$ and $2\mu$, for reference, with integer and half-integer $\spin$ saturating for $\hbar\omega>\mu$ and $\hbar\omega>2\mu$, respectively.} \label{fig:higher_S}
\end{figure}

In this appendix, we comment on the case of general pseudospin $\spin>2$.
As discussed in Sec.~\ref{sec:untilted},
D\'ora et al. \cite{doraLatticeGeneralizationDirac2011a} have calculated the optical conductivity of an untilted 2D Hamiltonian with general pseudospin $\spin$ and the result is summarized by Eqs.~(\ref{eq:2D_inter_notilt}) and (\ref{eq:2D_intra_notilt}). In Fig. \ref{fig:higher_S}, we plot the interband result for pseudospin up to $\spin = 7/2$. All integer values of $\spin$ will display a step at $\hbar\omega=\mu$ due to a flat-band-to-cone transition involving the $\lambda=1$ cone, while for all half-integer $\spin$ cases, $\lambda=1/2$ cone-cone transitions give rise to a step at $2\mu$. It is clear from this figure that for large values of $\spin$, the primary modification is at lower energies $\hbar\omega<\mu$, where
new steps in the optical conductivity spectrum appear at $\hbar\omega = \mu/\lambda$, with the effect of a `bunching up' of steps at lower energy as $\spin$ increases. Each additional step is due to the opening of a new set of transitions occurring between neighboring bands with higher $\lambda$ label. As the model Hamiltonian discussed in this work will only apply as a low energy theory,  and $\mu$, as discussed here, would then be an energy scale smaller than the cutoff energy of the theory, we expect that these additional steps might be quite close in energy.
Consequently, in any experiment involving real materials, these steps may be difficult to resolve due to additional broadening introduced by finite temperature and impurity scattering. Furthermore, as we have seen in Sec.~\ref{sec:2Dtilted}, tilting introduces more energy scales about these steps along with a broadened conductivity curve related to the tilting parameter. Moreover, when 3D is included (Sec.~\ref{sec:3Dtilted}), the linear background further suppresses these features. Hence, we expect that even if such an experimental system could be found or engineered to have higher pseudospin beyond $\spin = 3/2$ or 2, it is less likely it could be verified via optical experiments. As a result, in this work, we have only provided explicit results and discussion for $\spin$ up to 2. From the cases we have examined, the pattern is clear for how additional bands for higher $\spin$ will contribute and this is summarized in Eqs.~(\ref{eq:2D_inter_notilt})-(\ref{eq:3D_intra_notilt}), for 2D and 3D untilted; Eq.~(\ref{eq:2D_fullinter}) and Eq.~(\ref{eq:3D_fullinter}) and the following equations for 2D and 3D with a tilt; and Eq.~(\ref{eq:full_intra}) and the following for the Drude weights for all cases. 

\section{Generic Form of $\mathcal{F}_i(\lambda,\omega)$\label{app:generic}}

In Secs.~\ref{sec:2Dtilted} and \ref{sec:3Dtilted}, we provided formulas for the conductivity written in terms of a quantity $\mathcal{F}_i(\lambda,\omega)$ for 2D and 3D, respectively. These forms for the case of $\spin=1/2$ ($\lambda=1/2$) have been presented and plotted in previous literature (subject to accounting for a slightly different statement of the parameters in the starting Hamiltonian) \cite{houEffectsSpatialDimensionality2023}. 
The optical conductivity results we present in this work are dependent on different values of $\lambda$ which control the (untilted) slope of the different cones in the band structure.
Hence, in this appendix, for completeness, we plot and discuss the generic $\omega$-dependent and $\lambda$-dependent weighting functions $\mathcal{F}_{\bot,\parallel}(\lambda,\omega)$ for the 2D and 3D tilted cases presented in Eqs.~(\ref{eq:inter_parallel_2_under})-(\ref{eq:inter_perp_2_crit}) and Eqs.~(\ref{eq:3D_para_undertilted})-(\ref{eq:3D_perp_critical}), respectively. The admixture of the various regimes shown here gives rise to the rich structure seen in the conductivity predicted for higher pseudospin $\spin$ systems.

\begin{figure*}[ht]
    \includegraphics[width=\textwidth]{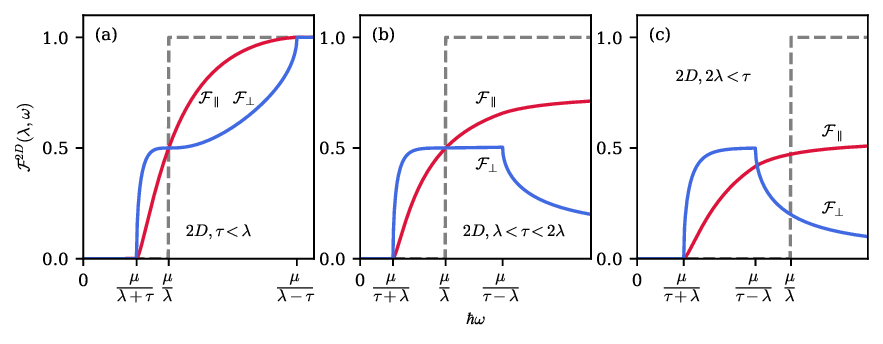}
    \caption{ The generic form of $\mathcal{F}^{\rm 2D}(\lambda,\omega)$ versus $\hbar\omega$ for general $\lambda$. Shown are three key regimes for the tilt parameter $\tilt$, compared to the untilted result (dashed curve): (a) $\tilt<\lambda$, (b) $\lambda<\tilt<2\lambda$, and (c) $\tilt>2\lambda$.}\label{fig:2D_generic}
\end{figure*}

Fig. \ref{fig:2D_generic} presents the generic behavior of $\mathcal{F}^\textrm{2D}_i(\lambda,\omega)$, where frame (a) reiterates the behavior of undertilted cones [Eq.~(\ref{eq:inter_perp_2_under})], while (b) and (c) both show the results for overtilted cones [Eqs.~(\ref{eq:inter_perp_2_under}) and (\ref{eq:inter_perp_2_over})].
For $\lambda<\tilt<2\lambda$ [Fig. \ref{fig:2D_generic}(b)],
the two conductivity polarizations are equal at $\hbar\omega = \mu/\lambda$, as found for the undertilted case [Fig. \ref{fig:2D_generic}(a)], but for $\tilt>2\lambda$ [Fig. \ref{fig:2D_generic}(c)] they do not cross at this point since the upper breakpoint is below $\lambda/\mu$ (\textit{i.e.} we begin to `lose' blue transitions before half of the red transitions are activated, see Fig.~\ref{fig:circles} discussion).
The critically tilted $\tilt = \lambda$ case (`type-III', not shown) can be understood as the limiting case of the regular overtilted behavior, where the upper breakpoint approaches infinity.

\begin{figure*}[ht]
    \includegraphics[width=\textwidth]{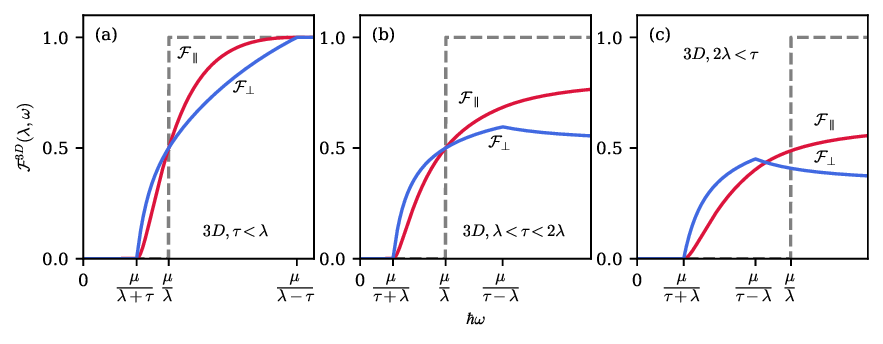}
    \caption{As in Fig. \ref{fig:2D_generic} but now for 3D, $\mathcal{F}^{\rm 3D}(\lambda,\omega)$ is plotted versus $\hbar\omega$. The final conductivity, based on this function, will have an additional factor of $\omega$ due to the dimensionality of the electronic density of states.}\label{fig:3D_generic}
\end{figure*}

For the 3D case, we have plotted $\mathcal{F}^\textrm{3D}_i(\lambda,\omega)$
in Fig. \ref{fig:3D_generic}.
Note that the linear background of the conductivity curves is not present here as that dependence enters as a prefactor to $\mathcal{F}^\textrm{3D}_i(\lambda,\omega)$ in the conductivity equation. Hence, Fig. \ref{fig:3D_generic} allows a
more interesting comparison with the 2D case and demonstrates that there is more than an overall $\omega$ factor affecting the 3D conductivity.
The two polarization directions are still less distinct here than in the 2D case (Fig. \ref{fig:2D_generic}) but similar momentum alignment-defined trends to those from 2D are present in both the under and overtilted cases (though the 4-dimensional energy-momentum space required to discuss these results in the same language as Fig. \ref{fig:circles} inhibits a similar visualization).

These plots of the $\mathcal{F}_i(\lambda,\omega)$ for general $\lambda$ in 2D and 3D can be used to visualize expected structures in the conductivity, resulting from the superimposed response of each $\lambda$-based component as it relates to $\tilt$. An example of the breakdown of this underlying structure was shown in Fig.~\ref{fig:perp_decomposed} of the main text.

\bibliography{tiltedSbib}

\end{document}